\documentclass{aa}  

\usepackage{graphicx}
\usepackage{txfonts}
\usepackage{txfonts}
\usepackage{xspace}
\usepackage{amsmath}
\usepackage{booktabs}
\usepackage{multirow}

\def\etal{{et~al.\null}}

\newcommand{\Ha}{H$\alpha$\xspace}
\newcommand{\HI}{\ion{H}{i}\xspace}
\newcommand{\HII}{\ion{H}{ii}\xspace}
\begin{document} 

   \title{Virgo Filaments VI: \Ha clumps in the filaments around the Virgo galaxy cluster}

   \author{G. Nagaraj\inst{1}\fnmsep\thanks{gautam.nagaraj@epfl.ch}
          \and
          P. Jablonka\inst{1,2}
          \and
          R.~A. Finn\inst{3}
          \and
          Y.~M. Bah\'{e}\inst{1,4}
          \and
          F. Combes\inst{5}
          \and
          G. Castignani\inst{6}
          \and
          B. Vulcani\inst{7}
          \and
          G. Rudnick\inst{8}
          \and
          D. Zakharova\inst{9}
          \and
          R.~A. Koopmann\inst{10}
          \and
          D. Zaritsky\inst{11}
          \and
          K. Conger\inst{8}
          }

   \institute{Laboratoire d'astrophysique, Ecole Polytechnique Federale de Lausanne (EPFL), Route de la Sorge, 1015 Lausanne, Switzerland
   \and
   LIRA, Observatoire de Paris, Universit\'e PSL, CNRS, 5 Place Jules Janssen, 92190 Meudon, France
   \and
   Department of Physics and Astronomy, Siena University, 515 Loudon Road, Loudonville, NY 12211, USA
   \and
   School of Physics and Astronomy, University of Nottingham, University Park, Nottingham NG7 2RD, UK
   \and
   Observatoire de Paris, LUX, Collège de France, CNRS, PSL University, Sorbonne University, 75014, Paris
   \and
   INAF—Osservatorio di Astrofisica e Scienza dello Spazio di Bologna, via Gobetti 93/3, I-40129, Bologna, Italy
   \and
   INAF—Osservatorio astronomico di Padova, Vicolo Osservatorio 5, I-35122 Padova, Italy
   \and
   University of Kansas, Department of Physics and Astronomy, 1251 Wescoe Hall Drive, Room 1082, Lawrence, KS 66049, USA
   \and
   INAF – Osservatorio Astronomico di Trieste, Via Tiepolo 11, I-34131 Trieste, Italy
   \and
   Department of Physics \& Astronomy, Union College, Schenectady, NY 12308, USA
   \and
   Steward Observatory, University of Arizona, 933 North Cherry Avenue, Tucson, AZ 85721-0065, USA
   }


    \abstract
    {It is still not clear which environmental processes operate in filaments. Given the ubiquity of filaments and their importance in feeding clusters, a proper understanding of these mechanisms is crucial to a more complete picture of galaxy evolution. To investigate them, we need large galaxy samples with spatially resolved information. As part of this effort, we analyse resolved \Ha maps of 685 galaxies inside and outside the filaments around the Virgo cluster in addition to extensive measurements of integrated physical properties. We create a pipeline to decompose the \Ha images into individual clumps that trace star forming regions. We find that the number and average size of clumps in a galaxy are well-defined functions of distance and angular resolution. In particular, the power-law relation between the number of clumps and the distance of a galaxy is consistent with a fractal structure of star forming regions. We formulate an algorithm to compare filament and non-filament galaxies after removing observational differences. Although we do not have any conclusive evidence for a difference in clump size distributions between filament and non-filament galaxies, we do find that filament galaxies have slightly more peripheral clumps than their non-filament counterparts.}


   \maketitle
%

\section{Introduction} \label{sec:intro}


Galaxy clusters have long been linked to increased quiescence \citep[e.g.,][]{DresslerGalClust1984,LewisEnvSF2002,WetzelGalEvoGroupClust2012,vandenburgclust2020} and early-type fraction \citep[e.g.,][]{DresslerMorphClust1980,PostmanGellerMorphDen1984,DresslerMorphDen1997,vulcaniclust2023}. Although these dense environments have been well studied, the vast majority of galaxies are found outside clusters \citep[e.g.,][]{CautunCosmicWeb2014,CuiEnv2019}. Surveys like SDSS \citep{York2000} and 2dF \citep{Colless20012dF} have helped form our picture of the large scale structure of the Universe as a cosmic web filled with long structures known as filaments \citep[e.g.,][]{BKP1996,KitauraCosmWeb2009,DarvishCosmWeb2014}. 

As a consequence, the research field exploring how galaxy evolution is driven or altered by the environment is steadily shifting from comparing galaxy properties in groups and clusters with those in the field — the definition of which is admittedly ill-defined — to considering the position of galaxies in the cosmic web, including the filaments that connect groups and clusters. 



Studies have shown higher early-type fractions \citep[e.g.,][]{KuutmaFilaments2017,CastignaniI2022,OkaneFilEff2024}, lower star-forming fractions \citep[e.g.,][]{CybulskiPreProc2014,BlueBirdFilaments2020,CastignaniI2022}, higher metallicities \citep{DonnanCosmicWeb2022}, and smaller star-forming disks \citep{CongerVirgo2025} in filament galaxies. However, it is not clear whether filaments enhance \citep[e.g.,][]{KleinerCosmicWeb2017} or repress \HI gas reserves \citep[e.g.,][]{OdekonFilaments2018, LuberLSS2019}. Furthermore, filaments are a heterogeneous environment and feature a complex intersection with groups. \cite{HoosainFilaments2024} and \cite{ZakharovaVirgo2024} find that while galaxies closer to filament spines are redder and more \HI-deficient, the dependence can be largely explained by enhanced group membership in the filament interiors. In addition, most environmental studies focus on quenched galaxies, so the impact of environment on star-forming galaxies is less understood. Overall, it is quite clear that a consensus on the environmental impact of filaments, let alone a full accounting of the underlying physical processes, is still distant.

The central question underpinning these investigations is that of how galaxies sustain and eventually quench their star-formation activity, also referred to as the baryon cycle. A series of seminal studies has revealed the connection between molecular gas and star-forming regions, thanks to the SINGS \citep{Kennicutt2003}, THINGS \citep{Walter2008}, and HERACLES \citep{Leroy2009}. \citet{Bigiel2008} found a linear relationship between SFR and H$_2$ surface densities ($\Sigma_{\rm{SFR}}$,  $\Sigma_{\rm{H_2}}$) at sub-kpc scales. Also, \citet{Leroy2008} showed that the H$_2$-to-\HI ratio , and by extension cloud formation, strongly depends on the galactocentric distance.

Built on that heritage, pushing spatial resolution down to $\le 100$pc scales, the PHANGS collaboration has shown that galaxies with higher stellar mass and more active star formation tend to host molecular gas with higher surface density and higher velocity dispersion \citep{Sun2020}. \citet{Rosolowsky2021} highlighted that the location of clouds is strongly influenced by the presence of stellar bars and spiral arms, providing the first sign of the relation between dynamical (local) environment and star formation conditions. The high-resolution, multi-wavelength journey of PHANGS continues, including PHANGS-MUSE \citep{Emsellem2022} and PHANGS-HST H$\alpha$ \citep{Chandar2025}, promising to provide constraints on feedback-regulated star formation.

There is as yet no comparable high-spatial-resolution study placing galaxies within their filament environment. One exception is the work of \citet{VulcaniFil2019}, who presented MUSE observations and spatially resolved properties of four nearby galaxies from the GASP sample \citep{PoggiantiGASP2017}, embedded in filaments identified by \cite{tempel2014}. These four galaxies present H$\alpha$ clouds beyond four times their effective radii and generally disturbed \Ha distributions, which is quite unusual outside a cluster environment. However, it is necessary to assemble a statistically representative sample that will allow us to assess the impact of filaments.




Thanks to its proximity, the Virgo cluster and its associated network of filaments offer a unique opportunity for detailed, spatially resolved studies. The work presented here is set within this context, as part of a series of papers that aims to better understand the full range of environmental conditions around Virgo. The first study, \cite{CastignaniI2022}, assembled \HI 21-cm and CO observations of 245 galaxies and carefully quantified a set of filaments motivated by but independent of \cite{KimVirgoFil2016} to understand the environmental impact of filaments on atomic and molecular gas. Compared with isolated galaxies from the AMIGA sample \citep{VerdesMontenegro2005} and Virgo cluster galaxies from \cite{BoselliCOHI2014}, they found a clear sequence from field to filament to cluster in terms of decreasing gas content and SFRs. The second study, \cite{CastignaniII2022}, put together a catalogue of nearly 7000 galaxies inside and around the Virgo cluster with an extensive collection of integrated data and basic physical properties (described in more detail in Section \ref{sec:obs}). Follow-up studies have included the comparison of the gas properties in filaments to results from a semi-analytic model \citep{ZakharovaVirgo2024}, an analysis of the effects of environment on the relative sizes of star-forming and stellar disks \citep{CongerVirgo2025}, and the detailed characterisation of the disruption of the baryon cycle in the group NGC 5363/5364 based on \Ha and \HI data \citep{FinnFil2025}. 

The \Ha data used here \citep{FinnFil2025} are part of a much larger effort to obtain resolved observations of 685 galaxies inside and near the Virgo filaments (Finn \etal~in prep). In this work, we use the full set of \Ha observations to quantify the morphology of star formation in galaxies inside and outside filaments. In particular, we identify clumps in the \Ha images: collections of \HII regions whose exact sizes depend on image quality and galaxy distance (see Section \ref{subsec:effects}).

The paper is organized as follows: In Section \ref{sec:obs}, we describe the observations used in this study, including the \Ha maps and the relevant data from the main Virgo catalogue. In Section \ref{sec:methods}, we detail the process used to quantify the \Ha clump distribution in galaxies, the steps taken to ensure a clean sample, and the algorithm developed to properly compare clumps in different populations of galaxies. We also discuss the effects of image resolution and galaxy distance on clump sizes. In Section \ref{sec:ip}, we detail the basic properties of filament and non-filament galaxies in our sample for context before moving to the results of the clump analysis in \ref{sec:res}. Then, we provide evidence for the fractal nature of \Ha clumps in Section \ref{sec:fractal}. Finally, we summarize our findings and suggest future directions in Section \ref{sec:conc}. In this work, we assume a Hubble constant of $H_0 = 74$ km s$^{-1}$ Mpc$^{-1}$ \citep{TullyMotion2008}.


\section{Parent Sample and Observations} \label{sec:obs}


\subsection{Sample selection}

Our filament galaxy sample is drawn from a catalogue of 6780 galaxies in the Virgo cluster and its surroundings, including multi-wavelength photometry, positional and environmental information (distance and coordinates, nearest filament, group membership, and local density), and basic integrated properties \citep[stellar mass, morphology, etc.,][]{CastignaniI2022,CastignaniII2022}. The galaxies in the catalogue are assembled using a base of HyperLEDA \citep{PaturelHyperleda2003, MakarovHyperleda2014}, the NASA Sloan Atlas \citep{BlantonSloan2011}, and the ALFALFA $\alpha$100 \citep{HaynesAlfalfa2018} surveys. The authors use redshift-independent distances from \cite{SteerNEDD2017} when available and compute distances based on details about the cosmic flow from \cite{MouldCosmicFlow2000} for the other galaxies.

In terms of environmental parameters, the authors use the position information of the galaxy sample to calculate the $5$-nearest neighbour densities ($n_5$), used as an estimate of local density on roughly megaparsec scales. The authors build on the filament identification by \cite{KimVirgoFil2016} to present a list of thirteen filaments around the Virgo cluster, providing parametric forms of their spines. \cite{CastignaniII2022} define galaxies as being inside a filament if they are within $2.70$ (2$h^{-1}$) Mpc from the filament spine. Group memberships are taken from \cite{KourkchiTullyGroup2017}. In this paper, we use the distances, local densities ($n_5$), stellar masses, Hubble types, \textit{r}-band elliptical fits (for galaxy size), morphologies, and the filament and group assignments.




We combine the information stated above with resolved \Ha observations of 685 galaxies inside or in the vicinity of Virgo filaments (from the entire sample of 6780). The observations are described below in Section \ref{subsec:obs}. In the process of observing filament galaxies, we have also been able to obtain \Ha imaging of many galaxies outside filaments. Approximately a third of our \Ha sample are outside identified filaments and can be used as a comparison sample. 

\begin{figure*}[!ht]
    \centering
    \resizebox{\hsize}{!}{
    \includegraphics{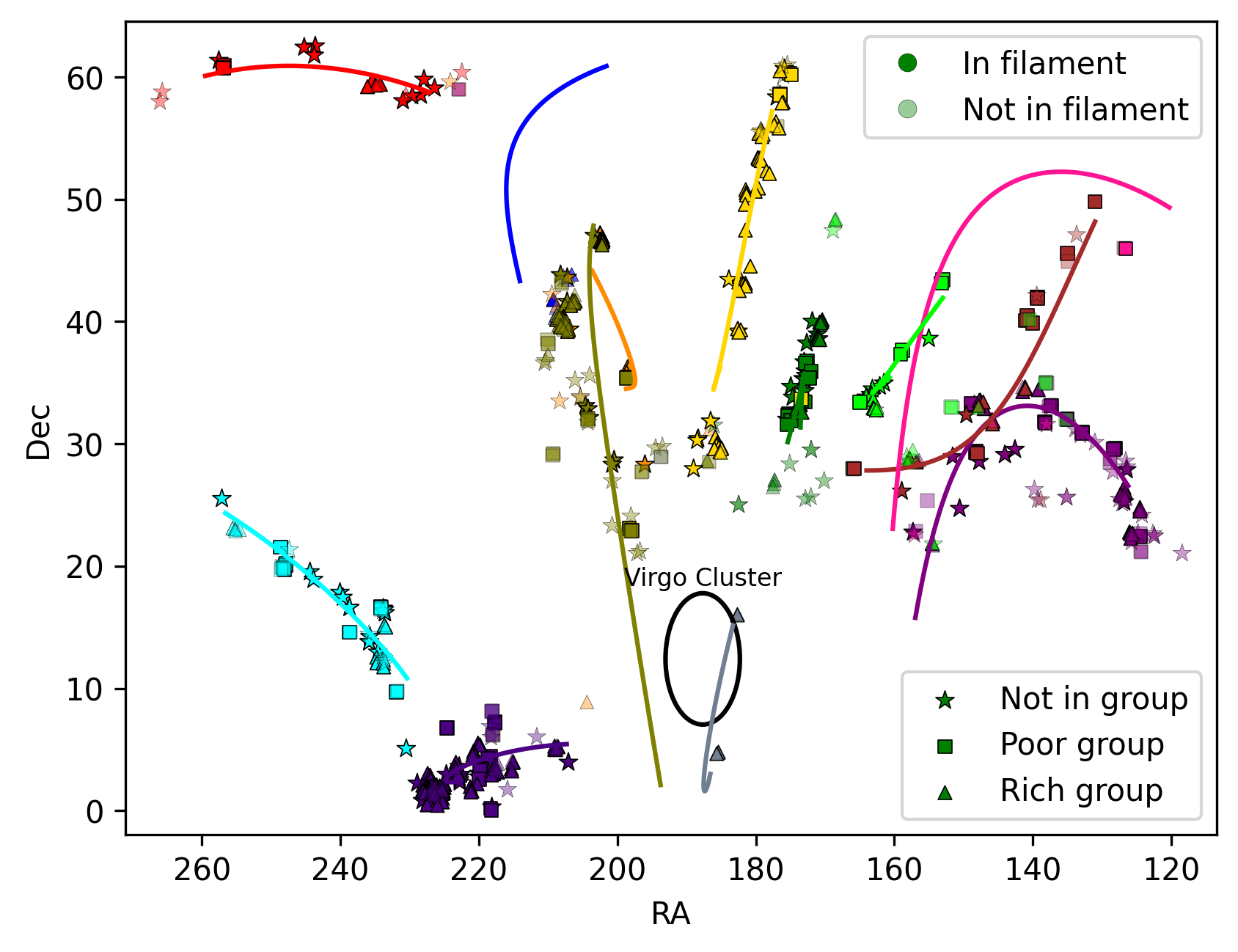}
    \includegraphics{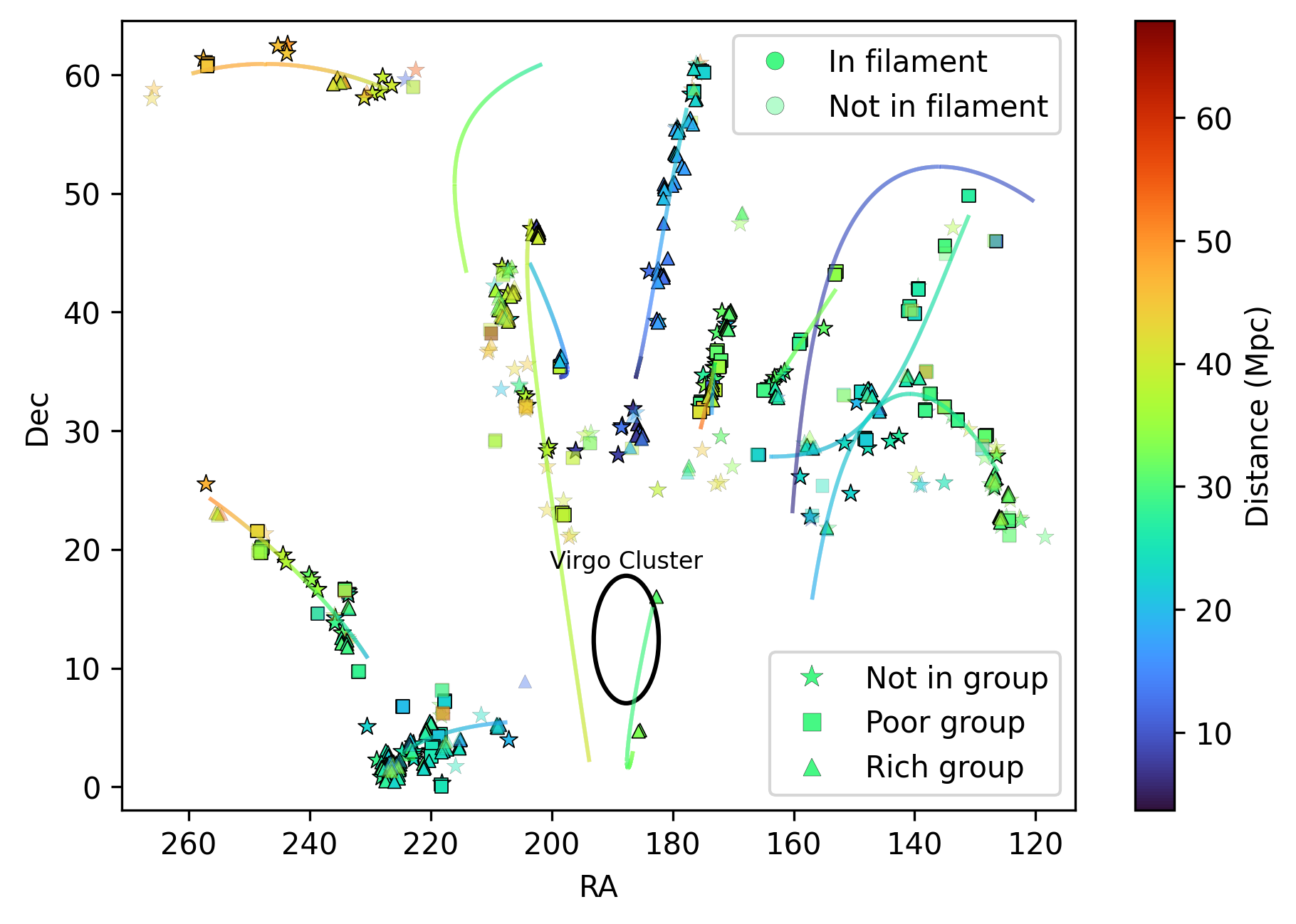}}
    \caption{Positions of all 685 galaxies in the \Ha sample. Galaxies represented by star symbols are not in galaxy groups, while square symbols are in poor groups and triangle symbols are in rich groups. Galaxies considered inside filaments (under $2.70$ Mpc from the spine of a given filament) have opaque symbols and thicker black edges while those outside the filament have translucent symbols and thinner black edges. We have indicated the Virgo cluster as a black circle. The radius corresponds to the virial radius of 1.55 Mpc \citep{McLaughlin1999}. In the left panel, we draw the 13 filaments studied in \cite{CastignaniII2022} in different colours and use the same colours to mark galaxies as being nearest to a given filament. In the right panel, we colour both the filaments and galaxies by line-of-sight distance to highlight the complex 3-D structure determining filament membership.}
    \label{fig:sample}
\end{figure*}

We show the spatial distribution of the 685 galaxies in the entire \Ha sample in Figure~\ref{fig:sample} along with their filament membership (in a filament or not) and group membership (rich group, poor group, or not in a group). The left panel shows which galaxies belong or are closest to which filament, while the right panel highlights the complex 3-D structure of the filaments and their constituents. Filaments have a mixture of galaxies in groups and galaxies not in groups, but there are several groups that are clearly centred around filament spines, showing the correlations between these environments. In the entire sample from \cite{CastignaniII2022} excluding cluster galaxies (for a total of 5628 galaxies), we find that 61\% of filament galaxies are in groups while only 43\% of non-filament galaxies are in groups, further highlighting this correlation.

%



\subsection{Observations} \label{subsec:obs}

Observations were conducted with narrow-band filters on four instruments: the Isaac Newton Telescope Wide Field Camera (hereafter referred to as INT), the Steward Observatory Bok Telescope 90Prime (BOK), the WIYN 0.9 m Mosaic 1.1 Camera (MOS), and the WIYN 0.9 m Half-Degree Imager (HDI). In all cases, the narrow-band images are accompanied by \textit{r}-band images. Details on the observations, including the filter descriptions and telescope details, are described by Finn \etal~(in prep). The data reduction and calibration procedures are reported in \citet{FinnFil2025}; we summarise the main points here. 

The images were reduced using standard methods to subtract the bias and dark current, and then flattened using either a dome or sky flat. We use \texttt{Scamp} \citep{Bertin2010} to solve for the astrometric calibration in each image, using the GAIA-EDR3 as the reference catalogue. The astrometric RMS error with respect to the GAIA-EDR3 reference positions is significantly less than a pixel and is typically 0.05\arcsec or less. We use \texttt{SWarp} \citep{Bertin2002} to coadd the images.  We determine the AB photometric zeropoint by comparing instrumental magnitudes with the reference \textit{r}-band magnitude from PAN-STARRS for both the \Ha and \textit{r}-band image. We create cutout images of each galaxy and make an accompanying mask that flag pixels not associated with the galaxy.  The masks are made by combining a segmentation image from Source Extractor \citep{BertinArnoutsSE1996} with known positions of GAIA-EDR3 stars.  When masking the stars, we adopt the relationship between stellar magnitude and star radius used by the Legacy Survey \citep[e.g.,][]{Zhou2023}. The masks are inspected individually to ensure that parts of the galaxy are not masked out.

The \Ha images contain continuum flux in addition to the line emission.  To remove the continuum, we follow the methods described in \citet{Kennicutt2008} and \citet{Boselli2018}. We use a scaled version of the $r$-band image to estimate the continuum in the \Ha image.  The scale factor depends on the difference in photometric zeropoints between the \Ha and $r$-band images and the color of the stellar continuum. We therefore include a term that scales with the $g-r$ color variations within each galaxy.  The $g-r$ color dependence is determined by integrating a large library of stellar spectra over the $r$ and \Ha filters (see \citealt{Boselli2018} for details).  
We construct a $g-r$ image for each target by reprojecting the Dark Energy Spectroscopic Instrument (DESI) {\it Legacy} Imaging Surveys images \citep{Dey2019} to match the FOV and pixel scale of the \Ha image.  The color-corrected continuum image is then subtracted from the \Ha image.

The full-width half maxima (FWHMs) of the point spread functions (PSFs) of the \Ha images range from 1.1\arcsec to 3.6\arcsec, with a median of 1.6\arcsec. Several galaxies (159 in our clean sample--see Section \ref{subsec:checks}) have multiple images because of poor weather and/or poor seeing conditions the first time. In our analyses, we consider only the best image (based on FWHM and visual inspection) for galaxies with multiple observations. 


\section{Methods} \label{sec:methods}

We quantify our galaxies' resolved star formation by identifying individual star-forming regions, or ``clumps.'' In Section \ref{subsec:clumps}, we detail our algorithm to divide the \Ha images into clumps, with the manual checks to ensure a clean galaxy sample described in Section \ref{subsec:checks}. There are several clump-finding methods in the literature, including \texttt{clumpfind} \citep{WdGBClumpfind1994}, source-extraction-based \citep[e.g.,][]{BertinArnoutsSE1996} software \citep[e.g.,][]{GuoClumps2015,MehtaClump2021}, and machine-learning algorithms \citep[e.g.,][]{Colombo2015,HuertasCompanyClump2020,AdamsClumps2025,Bazzi2025}. Our method uses a combination of wavelet analysis and watershed-type deblending. Wavelet analysis is useful for decomposing an image into different physical scales, allowing us to efficiently identify clumps.

\subsection{Analysis Pipeline} \label{subsec:clumps}

\begin{figure*}[!ht]
    \centering
    \includegraphics[width=0.86\linewidth]{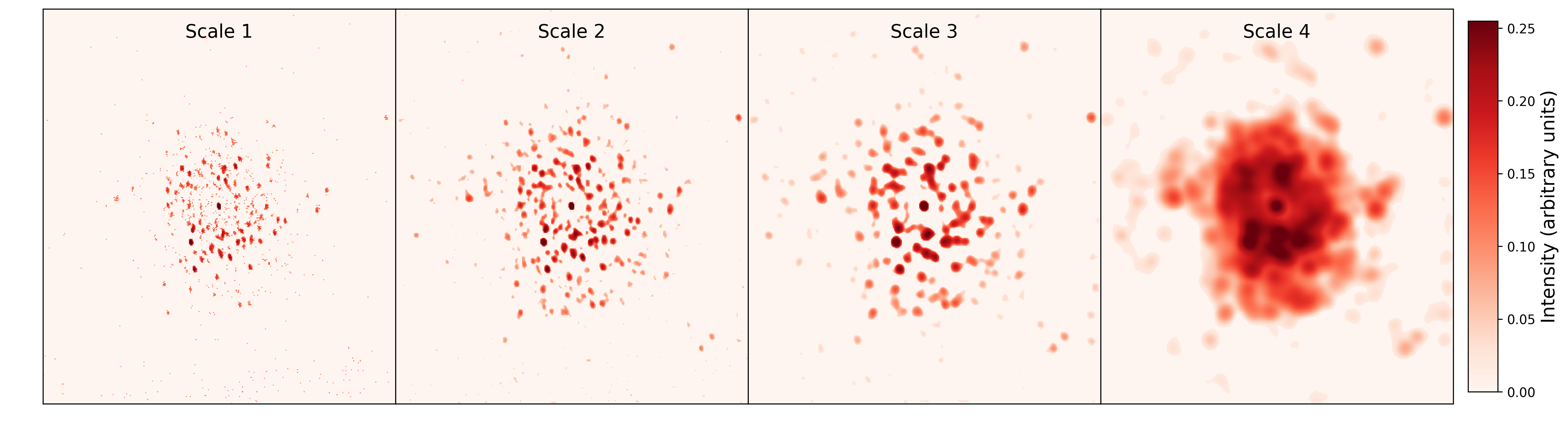}
    \includegraphics[width=0.48\linewidth]{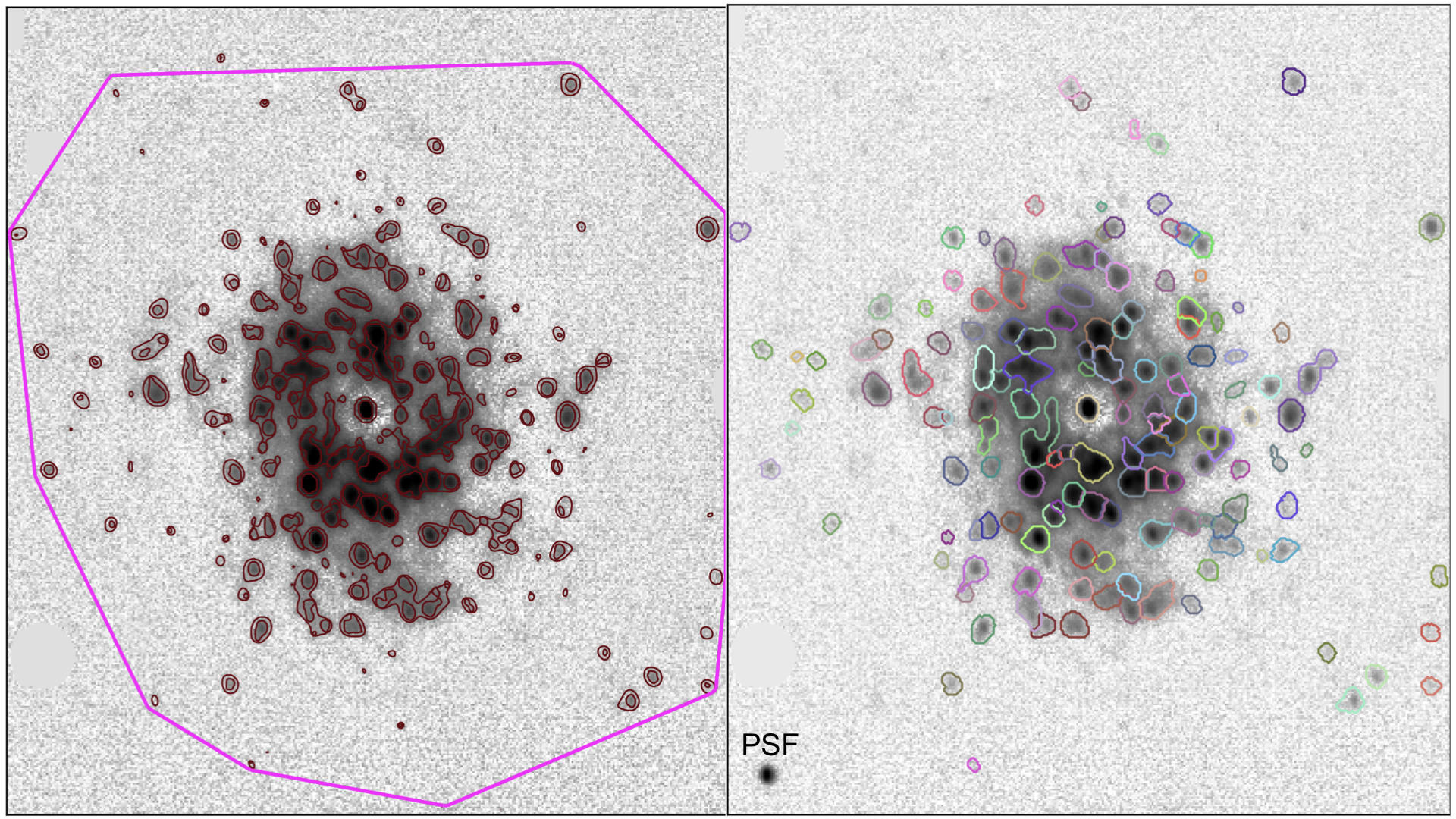}
    \caption{Illustration of the analysis pipeline for galaxy VFID 1035 (NGC 3982). In the top panel, we show the decomposition of the \Ha image into four Scarlet scales. In the bottom panels, the \Ha image is shown in grayscale (in arcsinh scaling). In the bottom left, we also show the Scarlet clump mask in burgundy contours, set to the 99.7th percentile of ($\sim 3\sigma$ above) the noise. Given the seeing FWHM of $1.39\arcsec$, this mask consists of both Scarlet scale two and three, visible as small contours within slightly larger contours. The galaxy boundary from the fourth Scarlet scale is marked in the fuchsia curve. In the bottom right panel, we show the final clump map after source detection and deblending by Photutils. In the lower left corner, we show the image PSF for size comparison; 95\% of the clumps are larger than the PSF. See the text for an extensive description of our analysis pipeline. }  
    \label{fig:methods1035}
\end{figure*}

To determine the clump structure of the 685 galaxies, we have created an analysis pipeline, illustrated by Figure \ref{fig:methods1035} for galaxy VFID 1035 (NGC 3982). First, we use wavelet analysis code Scarlet\footnote{https://pmelchior.github.io/scarlet/} \citep{MelchiorScarlet2018} to define clumpy SF regions in each galaxy. Wavelet analysis is the representation of functions as a particular family of orthonormal series (similar to the Fourier series), with applications for image compression and signal processing. In our case, wavelets enable the identification of features at different physical scales by decomposing the image into various spatial frequencies. In particular, we decompose galaxies into four scales. The first scale corresponds to emission peaks, always smaller than the PSF. The middle scales (second and third) identify regions of enhanced emission within the galaxy. The final scale encloses the entire galaxy, which is useful for marking the entire footprint. We show the decomposition of VFID 1035 into four scales in the top panel of Figure \ref{fig:methods1035}.

Several empirical tests we have performed show that the third scale corresponds to physical clumps. For galaxies seeing FWHM $\lesssim 1.8\arcsec$ (70\% of the sample), the second scale identifies smaller, more isolated clumps that are sometimes missed by the third scale. Therefore, our clump mask is defined as the entire region covered by the third scale as well as the second scale for galaxies with FWHM $\lesssim 1.8\arcsec$. When we quantify clumps (see below), we ignore emission outside this clump mask. In the bottom left panel of Figure \ref{fig:methods1035}, we plot the Scarlet clump mask (burgundy contours set at a threshold described later in the pipeline) on top of the VFID 1035 \Ha image (grayscale, arcsinh scaling). Given this galaxy's seeing FWHM of 1.39\arcsec, the clump mask consists of both scale two and three, visible as small contours within slightly larger ones.

The fourth scale is used to identify the boundaries of each galaxy. This is necessary to establish the limits of the clump mask, avoiding situations where we include extraneous emission that was either not properly masked out or caused by image artefacts. In almost all cases, we find (empirically) that including all pixels at or above the 16th percentile of the fourth-Scarlet-scale map perfectly envelopes the galaxy. Out of our 685 galaxies, we need to manually modify this threshold for 25, with percentile values ranging from 30-70. These are images with particularly widespread, low-level diffuse emission (possibly improperly subtracted continuum or brighter dust-scattered \Ha haloes, e.g., \citeauthor{SeonWittDiffuseHa} \citeyear{SeonWittDiffuseHa}), where the regions bounded by the 16th percentile contours are much larger than the galaxies. In the bottom left panel of Figure \ref{fig:methods1035}, we show the galaxy boundary for VFID 1035 (16th percentile contour of the fourth scale) as a fuchsia curve.

Although we now have a clump mask that is constrained to the galaxy footprint, Scarlet does not separate it into individual clumps. To identify the positions, sizes, and shapes of individual clumps, we use the Photutils package \citep{BradleyPhotutils2023}. First, we use their source detection algorithm on the clump mask, which is essentially a division of the Scarlet map into discrete units, with a few caveats. A region of an image is considered a clump if it contains at least eight contiguous pixels. In addition, we require that at least 5\% of the pixels in a clump are above a galaxy-specific threshold that is defined as follows. We identify the image background as all regions outside the boundaries set using the fourth Scarlet scale. We then define the 99.7th percentile of the \Ha ``emission'' (noise) in this background map to be the threshold, which is very similar to setting a 3-sigma detection threshold but is more stable against outliers. We find empirically that the 5\% criterion allows us to retain low-surface-brightness features while avoiding clumps without emission that are occasionally found by wavelet decomposition.

Finally, we allow for clump deblending. The Photutils package uses a watershed algorithm to divide \Ha emission regions into individual clumps. We set the following parameters for deblending, chosen through extensive testing to create clump maps consistent with detailed visual inspection: First, the minimum fraction of the total flux a component of a clumpy region must contain is 1\%. Next, the flux map across the region is divided into 32 linearly-spaced levels for finding saddle points to separate components. In addition, pixels are considered contiguous when sharing either an edge or a corner. Finally, we set the minimum number of contiguous pixels for a region to be divided into individual clumps as a number between 8 and 20 depending on the image FWHM (linearly scaled). 

In the bottom right panel of Figure \ref{fig:methods1035}, we plot the final clump selection of VFID 1035 as coloured curves on top of the \Ha image (grayscale, arcsinh scaling once again). A comparison of the individual clumps to the contours in the bottom left panel quickly demonstrates the conservation of the Scarlet features, though with compound structures typically deblended into multiple clumps. In the lower left corner, we show the PSF for size comparison. The majority of the clumps (95\% for VFID 1035 and 91\% for the entire sample of galaxies) are larger than the PSF (size calculated as $\pi (\rm FWHM/2)^2$). Only one percent of the clumps in the entire sample are less than half the PSF area. As Scarlet does not care about the PSF in this particular application (decomposition into spatial scales) and finds often very asymmetric features, it would be very difficult and even undesirable (given the added biases) to limit the clump-finding algorithm to those larger than the PSF. More importantly, we design stringent algorithms to compare populations of galaxies that take differences in the PSF into account (see Section \ref{subsec:fwhm}).

\subsection{Visual Checks} \label{subsec:checks}

To ensure a clean sample of \Ha clumps, we visually inspected each galaxy, in which we assigned a general use flag of 0 (unusable) to 2 (good). A use flag of 2 corresponds to galaxies with a clear \Ha signal, typically with a more clumpy structure than the continuum map, suggesting proper continuum subtraction. Galaxies with this flag value do not suffer from any artefacts. Examples of a galaxy that received a 2 in our rating system include VFID 1035, presented in Figure \ref{fig:methods1035}. The use flag of 1 is given to questionable sources, sometimes with large interloper galaxies partly or fully within the region of the main galaxy that could not be removed or sources with very little \Ha whose veracity is difficult to judge. Finally, the use flag of 0 is for catastrophic artefacts, complete contamination by a much brighter source, or a complete lack of \Ha. We use only sources with a use flag of 2.



We created other flags to indicate issues of masking of contaminants (foreground stars and other galaxies) covering parts of the galaxy as well as very bright central regions for which the clump-finding algorithm struggled. In addition, we used AGN classification from the HyperLEDA catalogue \citep{PaturelHyperleda2003, MakarovHyperleda2014} as assembled by \cite{CastignaniII2022} to create an AGN flag. We find no major impacts of the flags on our results. In any case, our final sample with a use flag of 2 and without AGN or galaxies with masking and/or bright-core issues contains 414 galaxies.

\subsection{Effects of Distance and Image Quality on Clump Properties} \label{subsec:effects}

An important question to consider is the physical significance of our measured clumps. Studies have shown that molecular clouds in the Milky Way and M31 have a median radius of $25$~pc and $22$~pc, respectively \citep[][based on hierarchical clustering algorithms]{MivilleDeschenesMCMW,ArmijosAbendano2025}. \HII regions in the Milky Way have radii ranging from $\sim 1-20$~pc \citep{TremblinHII2014}, with smaller sizes reflecting fewer stars and/or generally higher pressures in a given system. On the other hand, \cite{Barnes2026} find a much larger range of $\sim 1-250$~pc for \HII regions in nearby galaxies, with larger values arising from physical resolution limits of the instruments, especially in cases where deblending is difficult (e.g., crowded circumnuclear environments). 


Observations of local star-forming galaxies, such as those in the SINGS sample \citep{Kennicuttsings2003}, yield star-forming regions with sizes of tens to hundreds of pc \citep[e.g.,][]{WisnioskiClump2012}, suggesting that ``clumps'' are coherent complexes of molecular clouds. In general, it is clear that resolution plays an important role in determining the sizes and other properties of star-forming regions \citep[e.g.,][]{LivermoreClump2012,CavaClumps2018}. \cite{Kollmeier2017} compare images of star-forming regions at different physical resolutions and show the loss of information on features like shocks and ionisation fronts at resolutions worse than 25 pc/pixel (see their Figure 9). It is only with recent surveys like PHANGS-MUSE \citep{Emsellem2022} and SIGNALS \citep{RousseauNepton2019} that we are able to probe individual \HII regions in the local Universe.

As our Virgo filament galaxies span a large range of distances (6-68 Mpc) and image qualities (PSF FWHMs of 1.1\arcsec to 3.6\arcsec), the physical resolutions achieved (i.e., FWHM in pc) span a range of $\sim 1.2$~dex, from $\sim 45-720$~pc. Therefore, we need to be able to quantify the evolution of measured clump sizes and numbers as a function of the distance and angular resolution.


To help us with this task, we have run our analysis pipeline on a set of 5 galaxies from the SINGS dataset \citep{Kennicuttsings2003}: NGC 2976 (distance of 3.8 Mpc), NGC 3938 (12.7 Mpc), NGC 0024 (8.0 Mpc), NGC 0337 (19.4 Mpc), and NGC 1512 (12.5 Mpc). Their \Ha images have FWHMs of 0.92-2.29\arcsec, resulting in physical resolutions of 17-141~pc. The usefulness of this small sample is threefold: 1) The SINGS galaxy NGC 2976, with a physical resolution of 17~pc, is able to resolve all the way down to large individual \HII regions, giving us a more physical anchor to quantify the meaning of our clumps. 2) By using a completely independent sample and ensuring consistency with our Virgo galaxy sample, we demonstrate the robustness of our methodology. 3) The sample spans a factor of $5.1$ in distance and $2.5$ in angular resolution, providing a diverse set of probed physical scales.

\begin{figure*}[!ht]
    \centering
    \resizebox{\hsize}{!}{
    \includegraphics{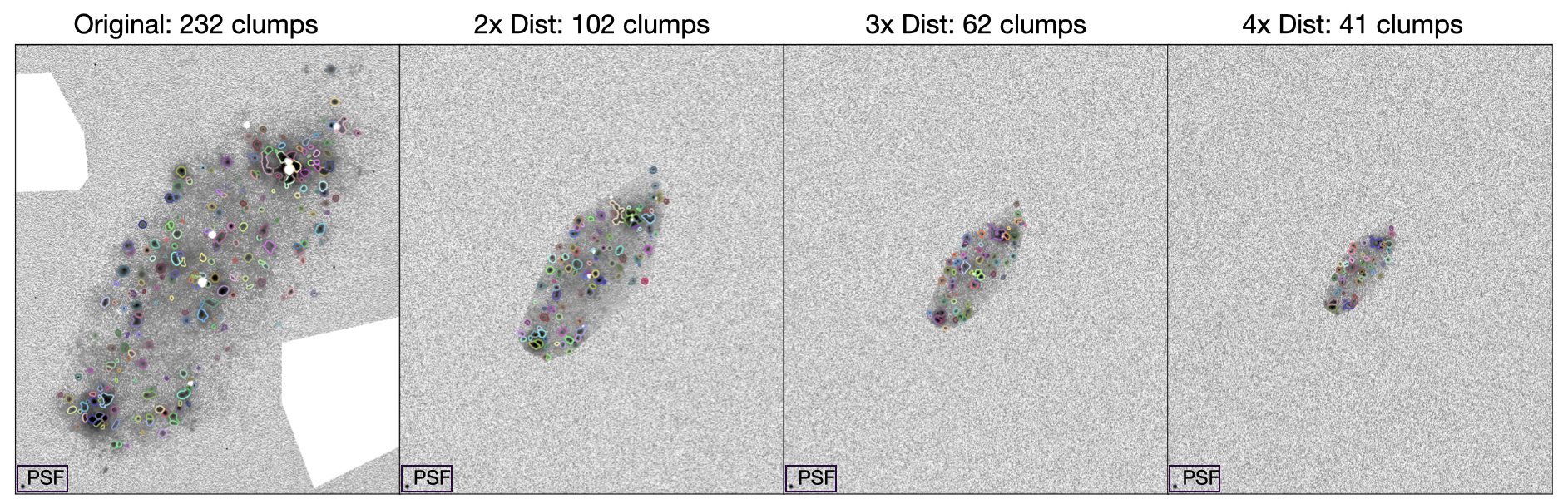}}
    \caption{Clumps detected by our analysis pipeline (coloured circles) in SINGS galaxy NGC 2976 at its original distance (first panel), and with the galaxy artificially moved to twice (second panel), thrice (third panel), and 4 times the distance (last panel, see text for details). We can see that while the clumps are always placed in the same general regions of the galaxy, the number of clumps decreases strongly with the distance.}
    \label{fig:clumpdistrel}
\end{figure*}

We conduct an experiment, shown in Figure \ref{fig:clumpdistrel} for NGC 2976, where we place the five SINGS galaxies at increasingly large distances (at factors of twice, thrice, and four times the distance, for example) without changing the angular resolution. The algorithm uses cubic spline interpolation (Scipy module \texttt{zoom}) and conserves surface brightness while shrinking the galaxy footprint so that the total flux is reduced by a factor of distance squared. 


Assuming a Gaussian noise profile, we fit a Gaussian curve (with mean fixed at 0) to all of the pixels outside the galaxy mask (based on the fourth Scarlet scale, see Section \ref{subsec:clumps}) in the original image. Thus, we obtain a value for $\sigma$ to quantify the noise profile. In the new image, everything outside the galaxy mask (which is now occupying fewer pixels) is populated with random values taken from the same Gaussian noise profile, thus making it so that we are mimicking the galaxy being at a larger distance but observed with the same telescope and night conditions (since background noise is distance-independent). We can see in Figure \ref{fig:clumpdistrel} that a couple of regions were masked out in the original image (leftmost panel) because of bright interlopers but disappeared in the modified images (other three panels) given the way our algorithm works. These regions have no impact on the measured clump distributions.

\begin{figure}[!ht]
    \centering
    \resizebox{\hsize}{!}{
    \includegraphics{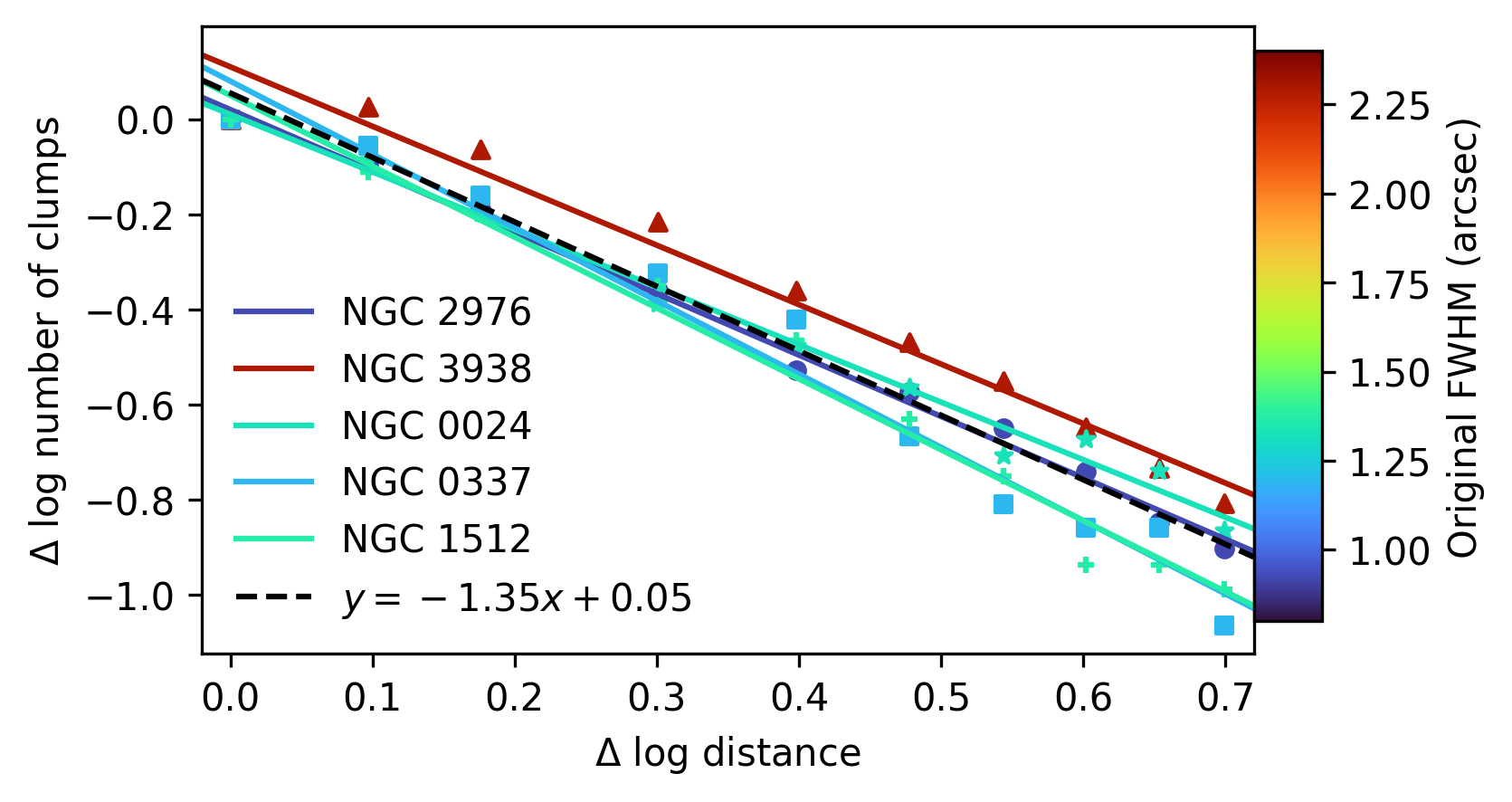}}
    \caption{Dependence of the number of clumps on distance for SINGS galaxies NGC 2976, NGC 3938, NGC 0024, NGC 0337, and NGC 1512, with lines coloured by the original seeing (FWHM) of the galaxy. The change in the number of clumps has a strong negative correlation (median $R^2=0.983$) with increased distance. When considering all five galaxies simultaneously, we find a slope of $-1.35 \pm 0.07$ (black dashed line). This suggests that \Ha clumps are hierarchical (fractal) in nature, with $D\sim 1.3-1.4$ (see Section \ref{sec:fractal}).}
    \label{fig:avgclump}
\end{figure}

It is clear that although the general regions of \Ha emission considered clumps are the same at all distances, the number of clumps decreases rapidly with increasing distance, with individual clumps generally representing larger physical areas in the galaxy. In Figure~\ref{fig:avgclump}, we show the change in the number of clumps measured as a function of distance for the five SINGS galaxies. The points (different markers) show the actual change in number of clumps in each experiment while the lines are best-fit curves through the points. We find that for each galaxy, there is a strong correlation between the decrease in number of clumps with the increase in distance (median $R^2=0.983$). In other words, although the behaviour of the changing number of clumps with distance is not exactly linear, the best-fit line is a very good approximation.

In Figure~\ref{fig:avgclump}, we colour the five SINGS galaxies (points and lines) by their angular resolutions (FWHM in arcsec). We find no clear trend between the slope or intercept and the FWHM, suggesting that the dependence of number of clumps on distance is independent of the FWHM. For all galaxies, we include the point $(0,0)$ as the original image and number of clumps, but we do not enforce a $y-$intercept of 0 in the fit given our interest in measuring an accurate slope. For this reason, the best-fit lines have intercepts close to zero but not exactly zero. The slopes vary between $-1.21$ (NGC 0024) and $-1.54$ (NGC 0337), but overall the spread in values is not large (standard deviation $\sigma=0.13$), indicating relatively uniform behaviour for a large variety of observing conditions (spanning a factor of $5.1$ in distance and $2.5$ in angular resolution). This tells us that the the number of clumps is a well-behaved function of distance whose relatively small scatter does not depend strongly on specific observational conditions.

A natural corollary of this conclusion is that the results are applicable to our Virgo filament galaxy sample as well. The application of the distance experiment to three Virgo filament galaxies (not shown in the paper) demonstrates this to be the case. Furthermore, the linear behaviour implies self-similarity, which we discuss in Section \ref{sec:fractal}. Specifically, we use the best-fit line for all five galaxies with slope of $-1.35 \pm 0.07$ to estimate the fractal dimension of \HII regions to be $D\sim 1.35$ in Section \ref{sec:fractal}. 



If we repeat the same test with the average clump size rather than number of clumps (not plotted in the paper), we find strong linear correlations (i.e., power laws), but with positive, and steeper slopes. With the same five galaxies, we find that the best-fit line has a slope of $1.79 \pm 0.07$ (and y-intercept of 0). If clumps at larger distances were simply combinations of clumps at smaller distances, we would expect a slope of $1.35$, equal to that of the number of clumps, but positive. The fact that we get a significantly steeper slope implies that clumps at larger distances also capture the low-density, low-surface-brightness regions between clumps identified at lower distances. We use this best-fit line to estimate observationally induced differences in clump sizes between populations with different median distances in Section \ref{subsec:clumpsmall}.



\begin{figure}
    \centering
    \resizebox{\hsize}{!}{
    \includegraphics{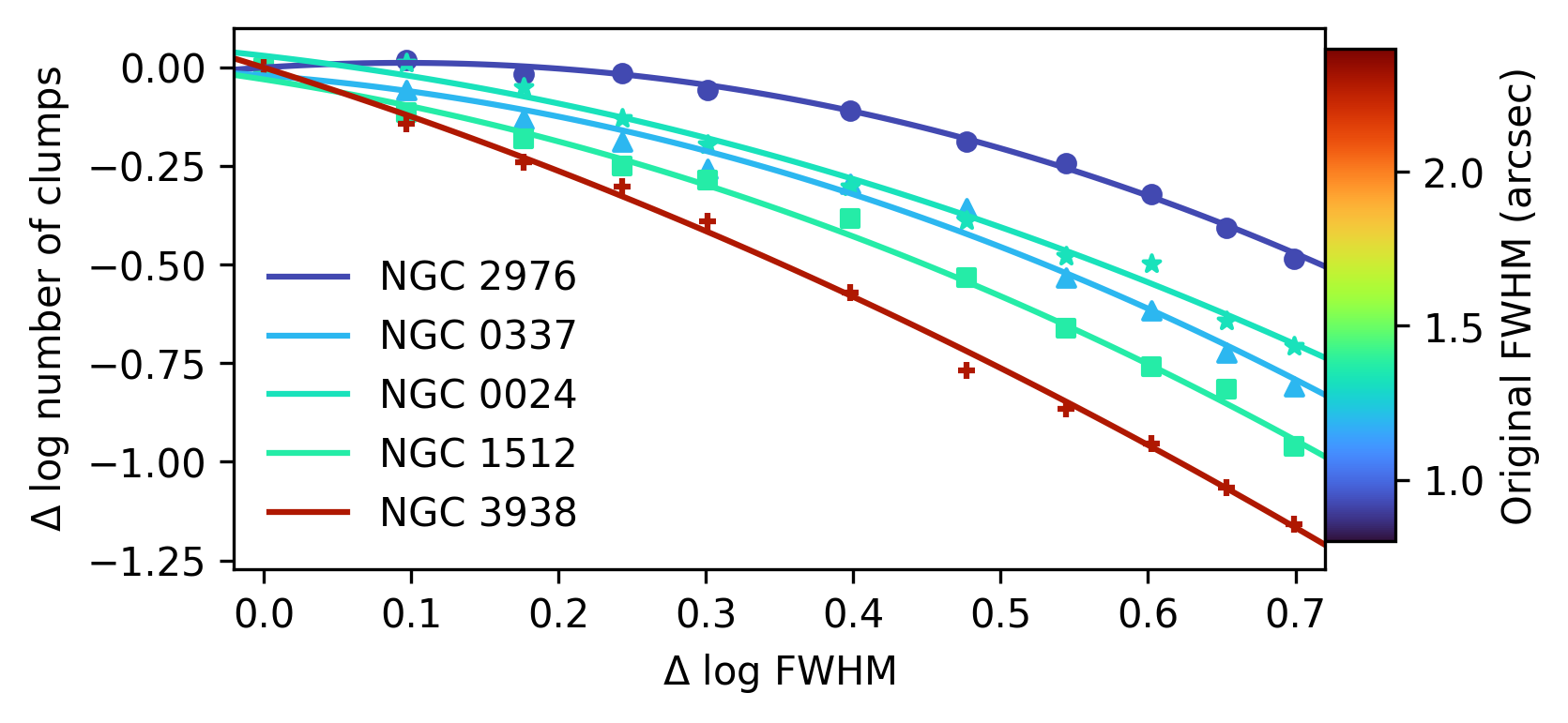}}
    \caption{Number of clumps as a function of angular resolution (FWHM) for the same five SINGS galaxies (NGC 2976, NGC 3938, NGC 0024, NGC 0337, and NGC 1512), with curves once again coloured by the original seeing (FWHM). The curves are well fitted with quadratic forms (median $R^2=0.995$) and have a large scatter that is significantly correlated with the original angular resolution, suggesting a more complex relationship with clump properties than distance, likely related to the angular scales associated with Scarlet scales two and three. }
    \label{fig:psfexpsings}
\end{figure}

The other important aspect affecting clump properties is angular resolution (image quality). In Figure \ref{fig:psfexpsings}, we conduct a similar experiment to that of Figures \ref{fig:clumpdistrel} and \ref{fig:avgclump}, but with PSF instead of distance. For the same five galaxies (NGC 2976, NGC 3938, NGC 0024, NGC 0337, and NGC 1512), we convolve the initial PSF (assumed to be Gaussian) with another Gaussian to make an effectively coarser PSF, up to factors $5\times$ the original FWHM. Specifically, $\Delta \log$~FWHM is a measure of the logarithmic ratio of the new FWHM (after convolution) to the original FWHM. We find that the number of clumps as a function of the increasing PSF is very well modelled by a quadratic function (median $R^2=0.995$), rather than a linear function, showing greater complexity with PSF than distance. 

Compared to the distance experiment, we find a much greater scatter between the five galaxies. There is a clear dependence of this scatter on the original angular resolution, though not monotonic: galaxies with higher resolution have a flatter response at small $\Delta \log$~FWHM (with the number of clumps nearly unchanged) while those with lower resolution immediately show a decrease in the number of clumps. We find the same trend but in the opposite direction for average clump sizes: galaxies with lower resolution have a flatter response at small $\Delta \log$~FWHM while those with lower resolution immediately start increasing in average clump size. 

The correlation between the shapes of the curves and the original FWHM is likely an indication of the effects of the Scarlet decomposition scales and Photutils deblending parameters we use: we are effectively probing a minimum angular size given our selection. In the case of NGC 2976 (dark blue curve), the angular resolution of $0.92$\arcsec, better than all galaxies in our sample (minimum resolution of $1.08$\arcsec), leads to a large flat response in the curve up to $\sim 0.25$~dex in $\Delta \log$~FWHM. This means we are almost exclusively probing features larger than the PSF in the original image. Given that the Scarlet scales and Photutils parameters were designed specifically for our Virgo filament sample, this is unsurprising and non-problematic. 

The quadratic curves and large variance in response due to differences in the original FWHM highlight the complexities of the effects of PSF on clump properties. Nevertheless, it is also clear that the response is well characterised. Furthermore, 94\% of our Virgo filament (and non-filament) sample have angular resolutions of FWHM$\leq 2.29\arcsec$, meaning with Figure \ref{fig:psfexpsings} and the corresponding version with average clump size, we are able to understand the effects of PSF on clump properties in our main sample. We performed the PSF experiment on a set of 20 Virgo galaxies and confirmed that the responses are also quadratic, with the same dependence on original FWHM. We explicitly use the average clump size relations with distance and FWHM to help understand our result in Section \ref{subsec:clumpsmall}. 

In this analysis, we have shown that the behaviour of clump sizes and numbers is a well-behaved function of distance and PSF. Furthermore, the effects of distance and PSF are comparable in magnitude: for a very large distance or FWHM increase of $5\times$ ($\Delta \log \rm{Quantity}=0.7$), the number of clumps is decreased by roughly one order of magnitude. Thus, when comparing clump distributions of two different populations, by carefully matching their physical FWHMs (which combine the effects of distance and angular resolution), we can minimise observational biases and learn true differences. We describe our matching algorithm in Section \ref{subsec:fwhm}.

\subsection{Creation of Comparable Populations} \label{subsec:fwhm}




\begin{figure*}[!ht]
    \centering
    \resizebox{0.92\hsize}{!}{
    \includegraphics{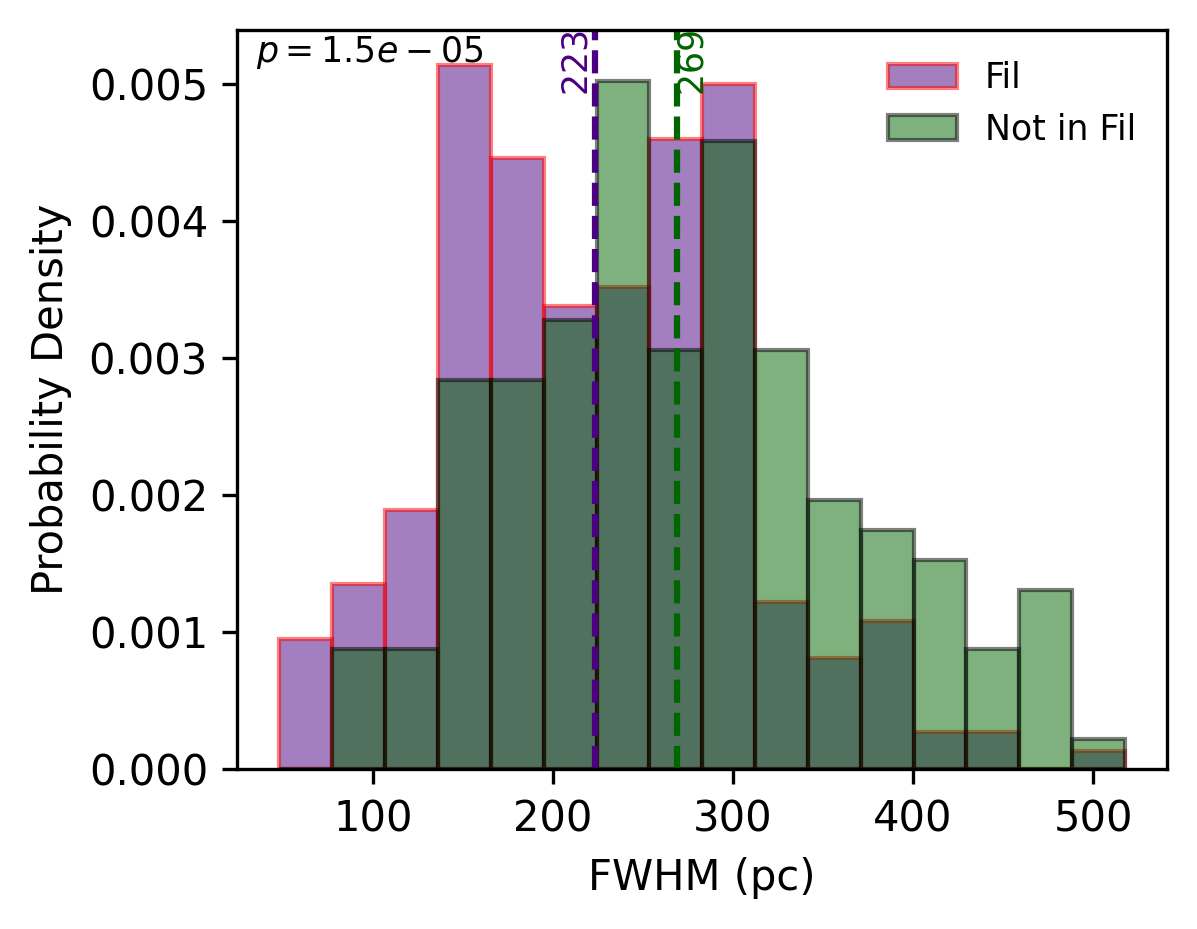}
    \includegraphics{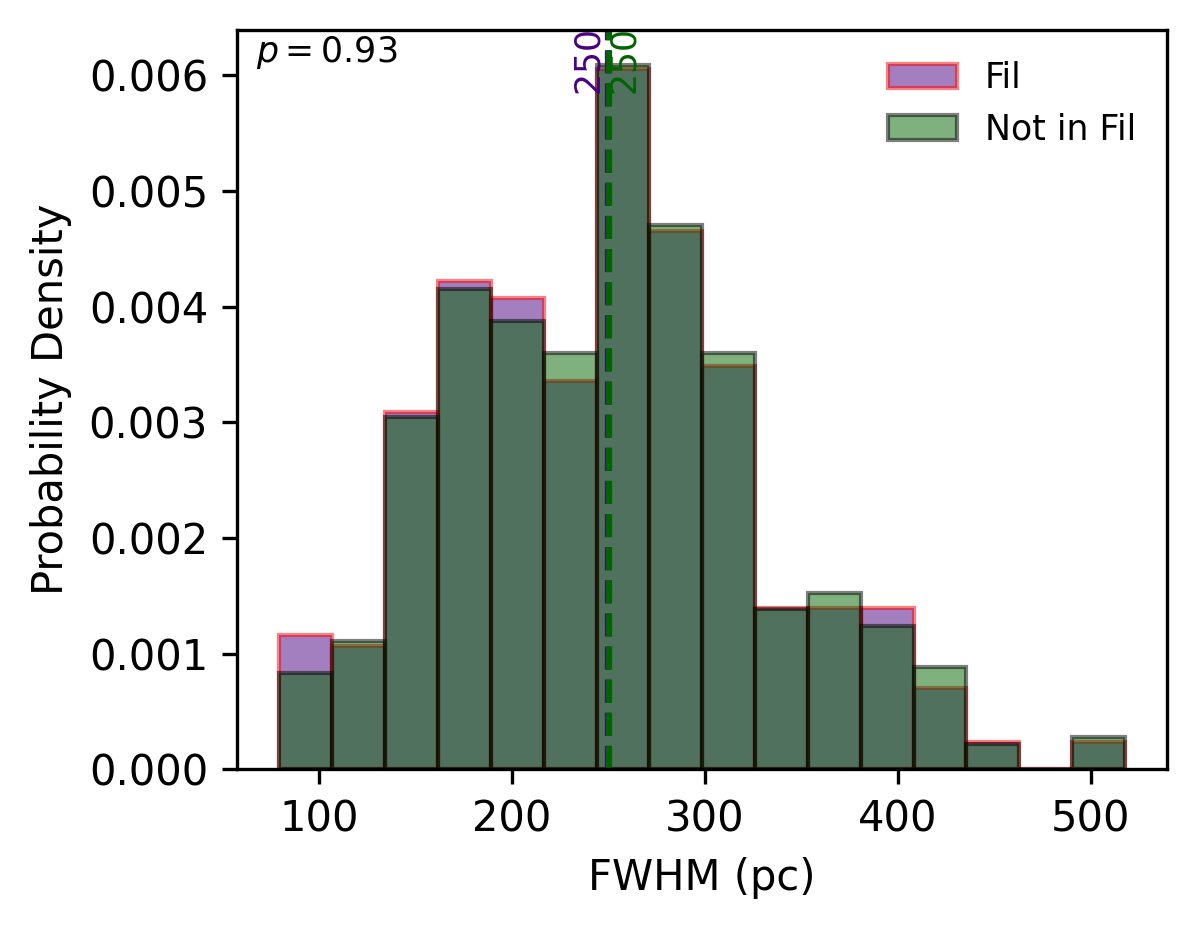}}
    \caption{Demonstration of the FWHM-compatibility procedure. In the left panel, we show the original sample of 252 filament (purple) and 162 non-filament (green) galaxies comprising the full set of 414 galaxies. After removing 95 filament and 29 non-filament galaxies, we are left with the distributions on the right, whose FWHM distributions follow each other very closely. In both panels, the median FWHM of each population is shown as a dotted line. The medians are identical after the procedure.}
    \label{fig:fwhmfil}
\end{figure*}

In Section \ref{subsec:effects}, we have shown that physical resolution (FWHM in pc) is essential in determining the meaning of individual clump sizes. As a result, any analysis aiming at understanding clump distributions involving diverse populations of galaxies must account for resolution effects. All of our resolved-clump results involve comparing the details about the clumps in filament vs.~non-filament galaxies. To ensure that the comparisons made are strictly related to the difference between the populations, we need to guarantee that their distributions of physical resolution are nearly identical. The precise details of this procedure are listed below.

\begin{enumerate}
    \item We calculate the FWHMs of all the galaxy images in physical units (i.e., parsecs) based on the FWHMs in pixel units, the pixel scales, and the measured galaxy distances. 
    \item We divide each galaxy population (of the ones being compared) into equally-spaced bins between the range of physical FWHMs occupied by both populations. Galaxies below or above the range are immediately removed. The number of bins is chosen as the floor of the size of the smaller population divided by 10. For example, with 252 filament and 162 non-filament galaxies, we select 16 bins. The number of bins was chosen empirically to enforce a strong population match while avoiding very sparsely populated bins.
    \item For a pair of two populations (i.e., filament vs.~non-filament), we equalise the histograms. In each bin, we compute the number of galaxies in that bin divided by the total number of galaxies for each population. Then, we randomly remove galaxies from the population with the larger fraction until the difference is minimised. 
    \item Since the total number of galaxies changes as we move through the bins, the fraction of all galaxies in any given bin also changes. To clarify, consider the following scenario: After going through bin 1 out of 16, we have 15 filament galaxies and 10 non-filament galaxies in the bin, and 240 filament galaxies and 160 non-filament galaxies in the entire sample (all 16 bins). Thus, the fraction of filament galaxies within bin 1 is $15/240=0.0625$, and the fraction of non-filament galaxies is $10/160=0.0625$, meaning the populations are equalised. After going through the other 15 bins, there were only 150 filament galaxies and 125 non-filament galaxies left in the sample, meaning the bin 1 filament fraction changed to $15/150=0.1$ while the non-filament fraction changed to $10/125=0.08$, pushing the populations out of equilibrium. Therefore, we need to repeat the process iteratively, with the ending condition that the difference in the fraction of total sources in each bin is the minimum possible value. For subsequent iterations, we also allow the possibility of adding back removed galaxies. This works in the same way as the normal procedure in step 3: in a given bin, if the population with the smaller fraction already has a list of removed galaxies, we add them back in random order until the difference is minimised.
    \item Steps 3-4 above remove or add back randomly-selected galaxies in each bin until the fractions are as equal as possible. Thus, we will get a different set of galaxies each time we conduct the procedure. To provide stable results, any time we do the matching process, we repeat it 25 times. 
    
\end{enumerate} 

To make this process more clear, we show the distribution of uncorrected (left panel) and corrected (right panel) FWHMs of filament vs non-filament galaxies considered in the study in Figure~\ref{fig:fwhmfil}. The full population of 414 galaxies is divided into 252 filament and 162 non-filament galaxies. Using the criteria explained above, we find that the two distributions are initially incompatible: a Kolmogorov-Smirnov (K-S) 2-sample (two-tailed) test yields $p=1.5 \times 10^{-5}$, suggesting that the distributions are distinct (which is obvious when viewing the figure). Our procedure to equalise the population results in the removal of 95 filament and 29 non-filament galaxies. The subsequent distributions are essentially identical ($p=0.93$).

\section{Integrated Properties of Filament and Non-Filament Galaxies} \label{sec:ip}

\begin{figure*}[!ht]
    \centering
    \resizebox{0.92\hsize}{!}{
    \includegraphics{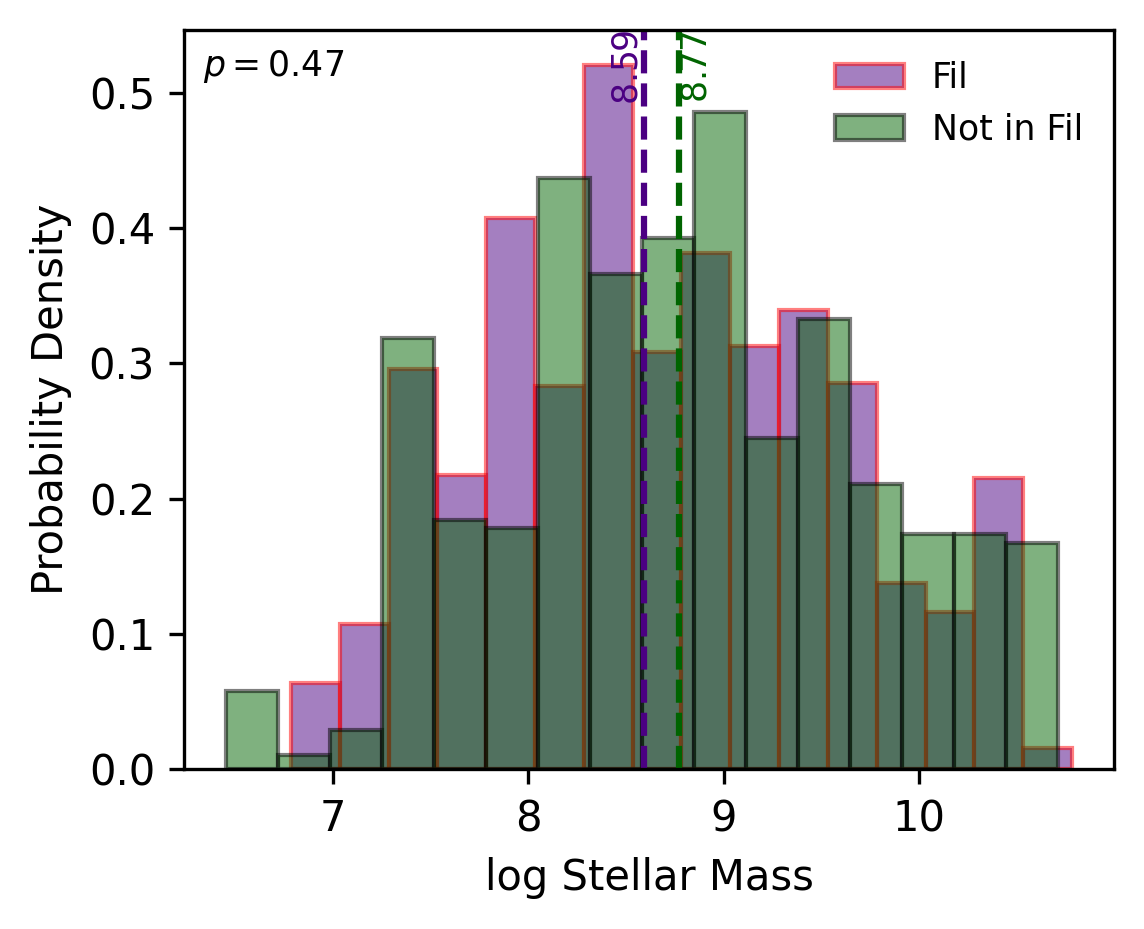}
    \includegraphics{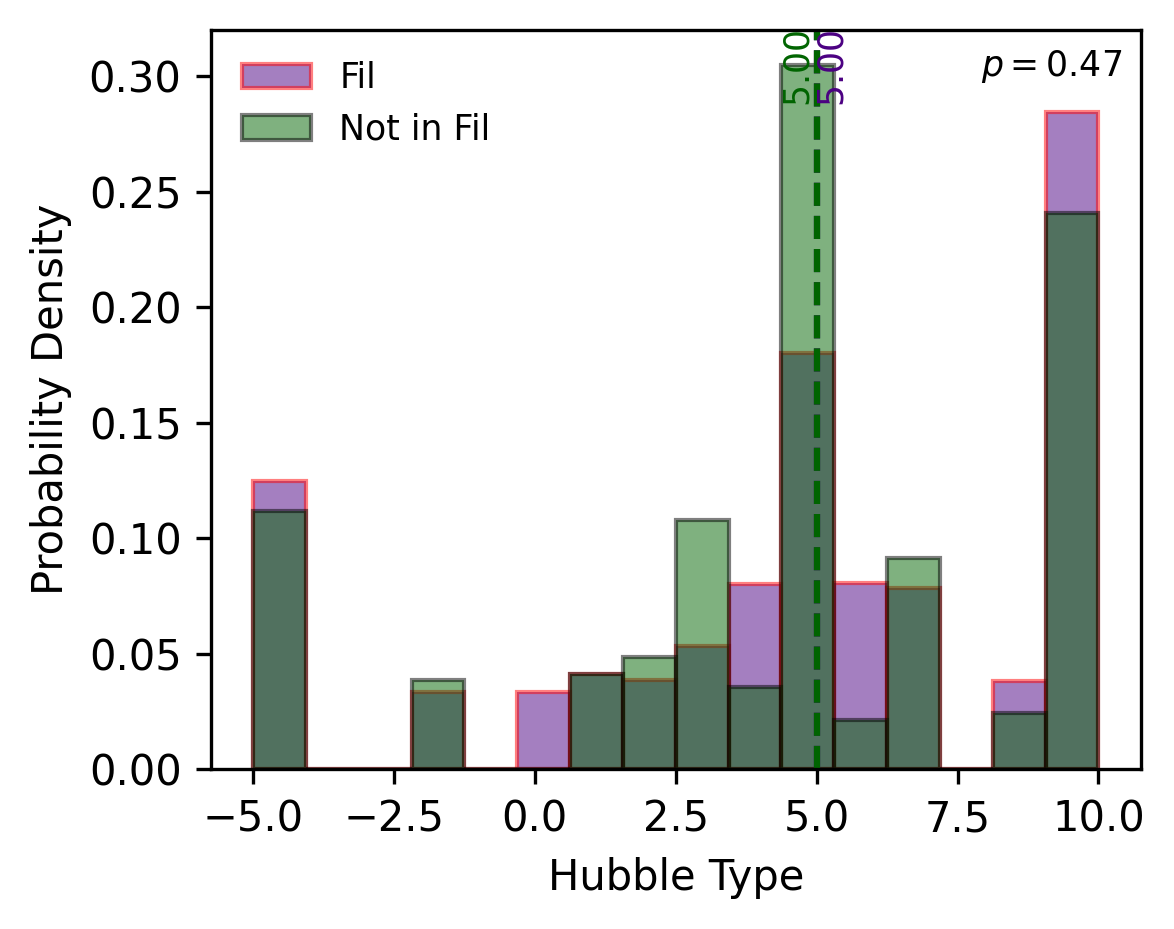}}
    \resizebox{0.92\hsize}{!}{
    \includegraphics{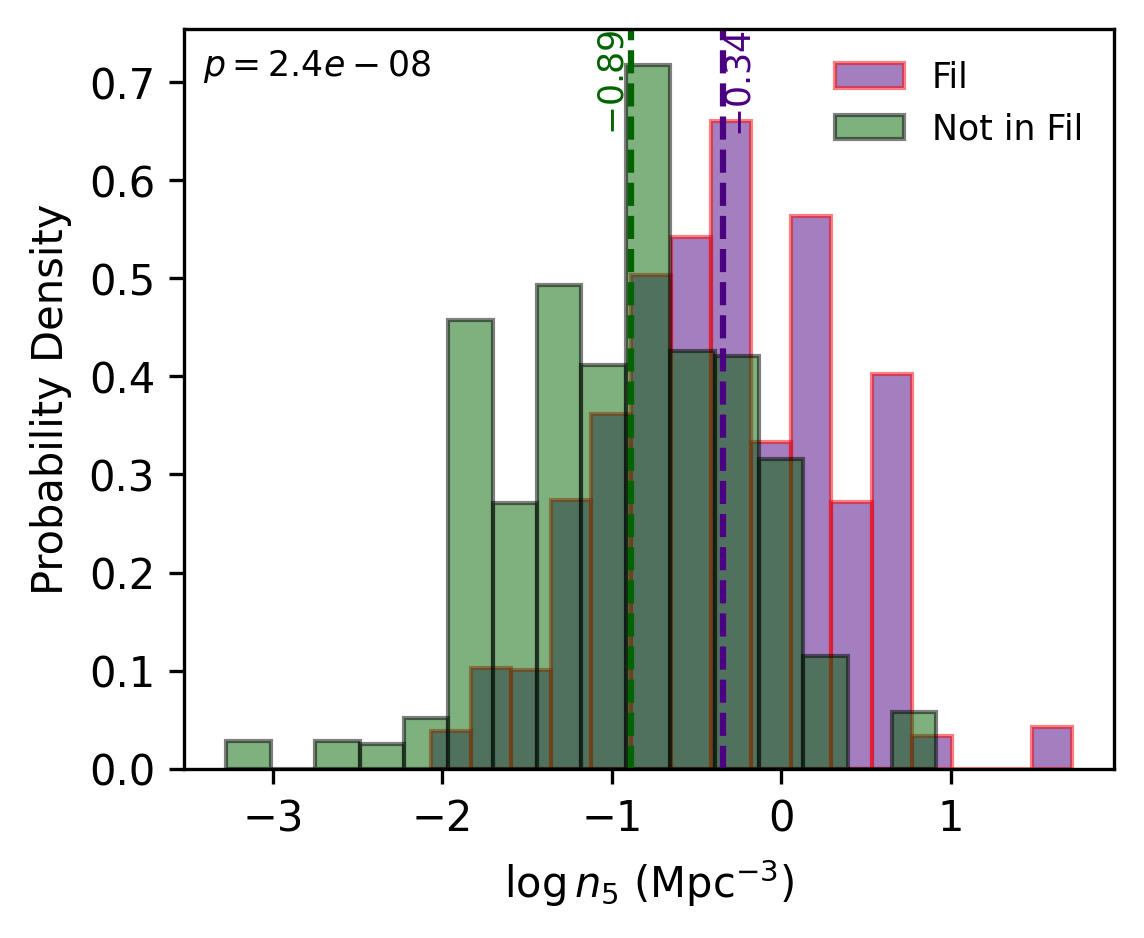}
    \includegraphics{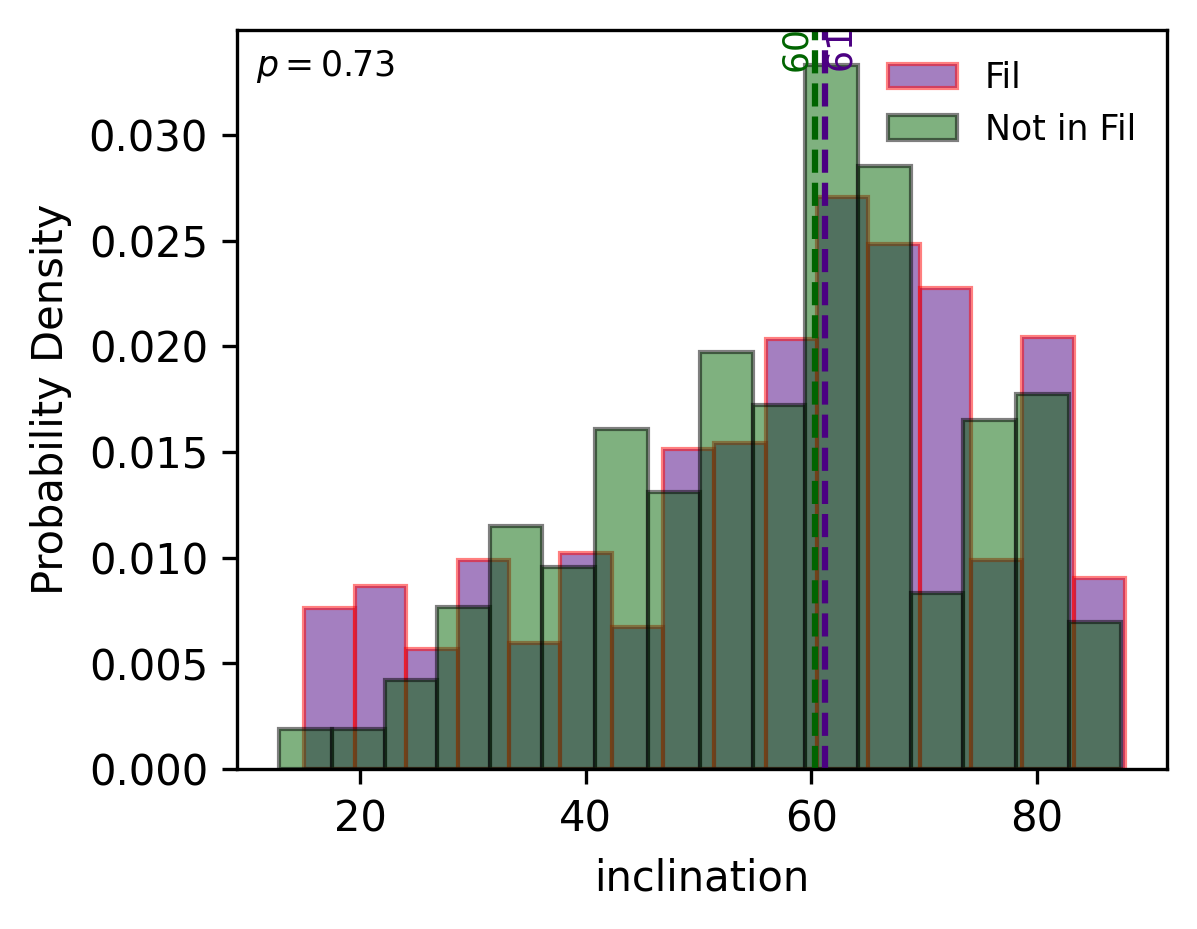}}
    \resizebox{0.92\hsize}{!}{
    \includegraphics{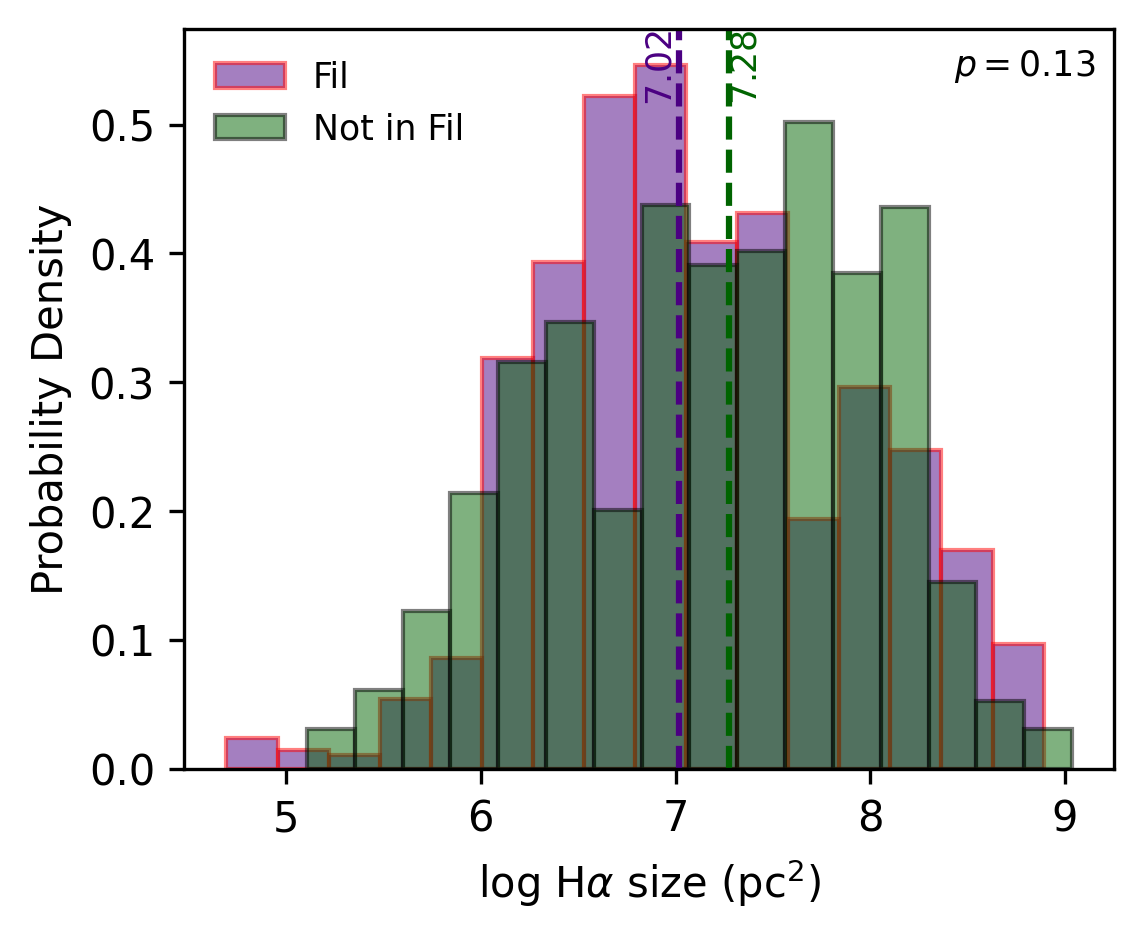}
    \includegraphics{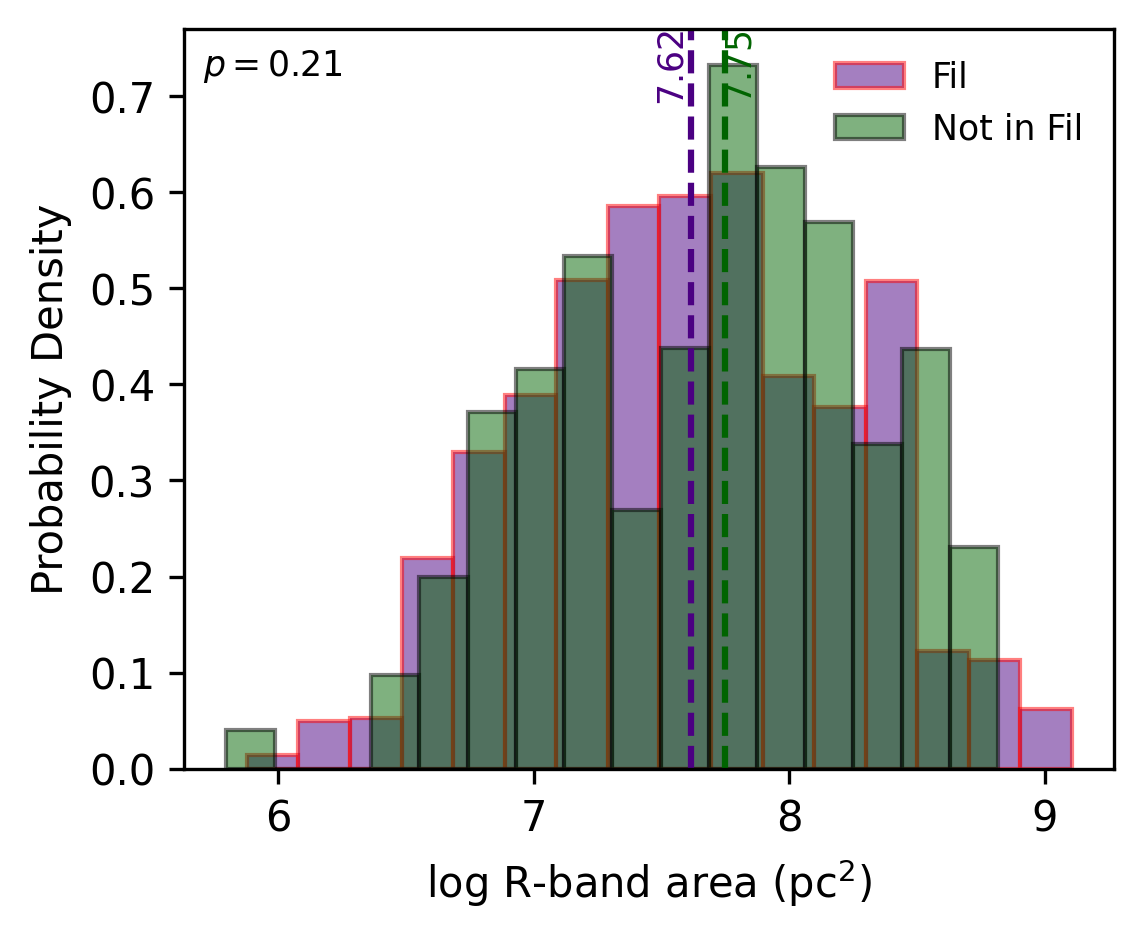}}
    \caption{Comparison of the distribution of various quantities for (PSF-matched) filament (purple) vs non-filament (green) galaxies in our sample (not mass-complete). We show stellar mass (top left), Hubble type (top right), local density $n_5$ (middle left), inclination (middle right; 90$^\circ$ refers to edge-on), \Ha size (bottom left), and \textit{r}-band size (bottom right). The median values of each distribution are marked as vertical dashed lines (purple for filament and green for non-filament). Our filament galaxies have much higher local densities and possibly smaller sizes than non-filament galaxies but are otherwise relatively similar in terms of physical properties.}
    \label{fig:filprops}
\end{figure*}

In previous sections, we have detailed the creation of a sample of 685 galaxies in and around filaments associated with the Virgo cluster; the detection and measurement of \Ha clumps; the manual checks to ensure purity, leading to a subsample of 414 galaxies; and the development of an algorithm to unbiasedly compare populations of galaxies. For our filament and non-filament comparison, the matched populations consist of 157 (after removing 95 of the original 262) and 133 (after removing 29 of the original 162) galaxies, respectively, for a final population of 290 galaxies (though in fact there are more than 290 galaxies represented since we always take 25 different iterations of the matching procedure). All subsequent results in the paper feature the matched populations.

In this section, we focus on basic (integrated) properties to provide context for the populations underlying the results in Section \ref{sec:res}. It is important to note that our samples are not mass-complete, as our goal is to highlight our methodology and compare clump distributions with as many galaxies as possible, and not to make definitive statements on integrated properties. Thus, the conclusions in this section are mostly limited to our sample. In Figure~\ref{fig:filprops}, we show the distributions of stellar mass, Hubble type, local number density $n_5$ (inverse of the volume occupied by the five nearest neighbouring galaxies), inclination, full \Ha size, and \textit{r}-band size for both filament (purple) and non-filament (green) populations. 

The \Ha sizes are defined by converting the total number of pixels within the fourth scale Scarlet map above the 16th percentile, or a different value for $\sim 3$\% of the population (Section \ref{subsec:clumps}), to a physical value in squared parsecs. All the other quantities are taken from the \cite{CastignaniII2022} catalogue. In the case of \textit{r}-band size, we define it by taking the moment-based elliptical fits (semi-major and semi-minor axes) derived by the 2020 Siena Galaxy Atlas \citep[SGA-2020;][available for the Virgo galaxies in the aforementioned catalogue]{MoustakasSGA2023} based on the Legacy Imaging Surveys Data Release 9 \citep{DeyLegacy2019} and calculating area as $A=\pi ab$.

Our galaxies span four orders of magnitude in stellar mass ($\sim10^{7}$ - $10^{11}$ $\rm{M}_\odot$) and include all morphological types from ellipticals to irregulars. The only property for which there is a strong statistically significant difference between filament galaxies and non-filament galaxies is $n_5$. Namely, non-filament galaxies tend to live in lower densities (0.54 dex) than their counterparts. This is simply a statement that the filaments are higher-density environments than non-filaments. 


Filament galaxies have a slight tendency to be smaller in \Ha and \textit{r}-band size than non-filament galaxies. The histogram also shows different peaks in the size distributions. These peaks (and the median difference) persist through a bootstrap analysis, suggesting they are real. At the same time, many of the largest galaxies in the entire sample are filament galaxies, showing the complexity of the comparison. In any case, the differences are not statistically significant ($\sim 1.5\sigma$ based on the p-value), meaning they are minor at most. We also note that the \Ha sizes are $\sim 0.5$~dex smaller than the \textit{r}-band sizes, but this is complicated to interpret given the difference in size definitions. 

Finally, there are galaxies at all inclinations, both with medians very close to the statistically expected 60$^\circ$ for randomly distributed galaxies. Both the sizes and especially observed positions of clumps depend on the inclination of the galaxy, so having statistically indistinguishable inclinations provides confidence for fair comparisons. In the case of clump positions, given the large influence of inclination, we divide galaxies into four inclination bins for a more reliable comparison (Section \ref{subsec:clumpdisp}). In general, we find that our filament and non-filament galaxy populations are statistically indistinguishable in all major properties except their environment, which means they are ideal for pinpointing the effects of the filament environment on \Ha clump properties.

\section{Resolved Star Formation in Filament Galaxies} \label{sec:res} 

We now focus on characterising the properties of the clumps. For every clump, we measure three properties: size, flux, and position. In this section, we compare the properties of clump sizes and positions in our filament and non-filament galaxies. As we lack information on dust attenuation, no definitive luminosity measurements can be made. Once again, for all results in this section, we use the FWHM-matching algorithm described in Section \ref{subsec:fwhm} to suppress observational biases. 



\begin{figure}[!ht]
    \centering
    \resizebox{0.92\hsize}{!}{
    \includegraphics{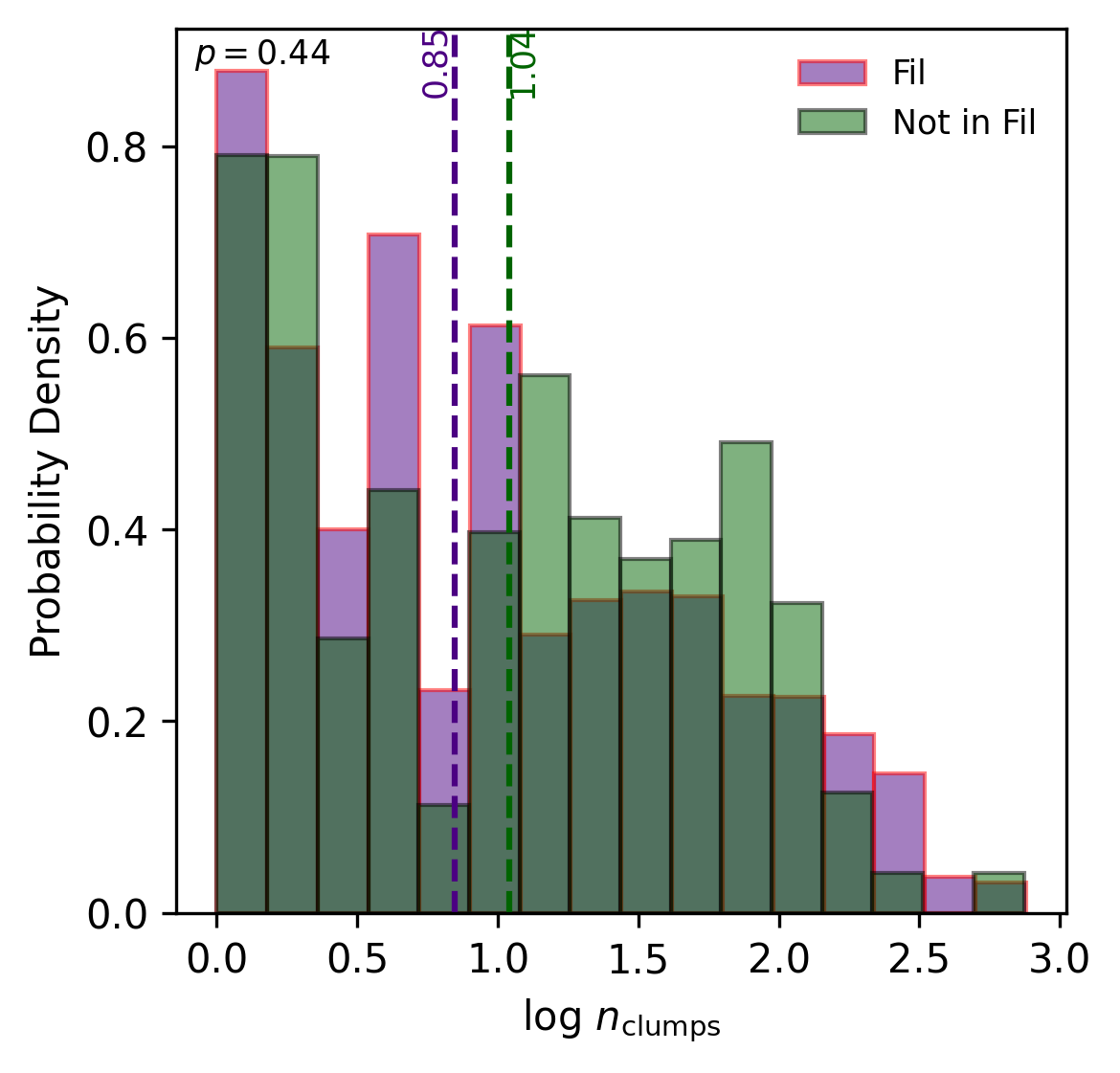}}
    \caption{Same as Figure \ref{fig:filprops} but for comparison of the number of clumps. Differences in the distributions are statistically insignificant ($p=0.35$). Nevertheless, the small differences we can see by eye reflect the differences between the \Ha size distributions in the bottom left panel of Figure \ref{fig:filprops}.}
    \label{fig:filncl}
\end{figure}

We first compare the total number of clumps for filament vs non-filament galaxies in Figure~\ref{fig:filncl}. According to a K-S 2-sample test, the distributions of the numbers of clumps are statistically indistinguishable ($p=0.34$, or a $1\sigma$ difference). The small differences we can see by eye---the lower filament median, and slight overabundances of filament galaxies with 1-12 and over 144 clumps---reflect the same shifts in the \Ha size distributions in Figure \ref{fig:filprops} (bottom left panel).

\subsection{Filament and non-filament galaxies generally have similar-sized clumps} \label{subsec:clumpsmall}

\begin{figure*}[!ht]
    \centering
    \resizebox{\hsize}{!}{
    \includegraphics{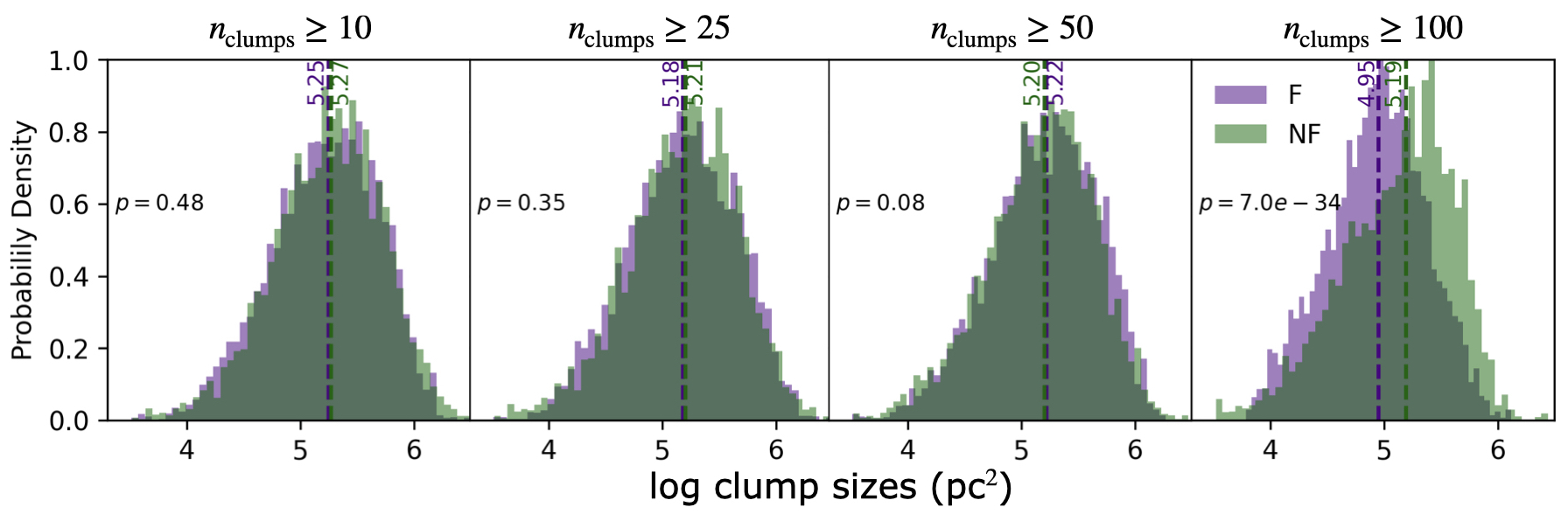}}
    \caption{The distribution of clump sizes in filament (purple) and non-filament (green) galaxies, with medians shown as dashed lines. In each panel, we take galaxies with at least $n$ clumps ($n$ indicated at the top) and randomly select $n$ clumps from the FWHM-matched populations (after the restriction of minimum number of clumps) to prevent biases toward either very clumpy (no weighting) or non-clumpy galaxies (galaxy-weighting). We find no differences in clump sizes until the $n=100$ case, where filament galaxies generally have smaller clumps than non-filament galaxies. This large difference may be caused by population differences arising from very small number statistics (only 10 non-filament galaxies). See text for a detailed discussion.}
    \label{fig:fil_sizeflux}
\end{figure*}

The clump size distribution is strongly dependent on galaxy distance and angular resolution, as shown in Figures \ref{fig:avgclump} and \ref{fig:psfexpsings}, respectively. Thus, with a galaxy sample whose physical FWHMs vary by over a factor of five, giving meaning to clump size is particularly difficult. Nevertheless, our careful PSF-matching procedure and the statistically indistinguishable stellar mass distributions of our samples allow us to compare the clump size distributions of our filament and non-filament galaxy populations. 

Another important consideration is how clumps from different galaxies are counted. A galaxy with 500 clumps provides 500 times more input than a galaxy with 1 clump in a simple clump stacking (i.e., aggregating the clumps of all galaxies in a given population for the analysis). On the other hand, we would strongly bias our results toward non-clumpy galaxies if the two galaxies were given equal weight, especially since galaxies with very few clumps tend to be farther away and have worse angular resolutions. To mitigate this issue, we randomly select a certain number of clumps from each galaxy. By choosing a few different numbers, we can get a more comprehensive and fair representation of the clump size distributions. 

In Figure \ref{fig:fil_sizeflux}, we compare clump sizes for filament (F; purple) and non-filament (NF; green) galaxies for four different numbers of clumps: 10, 25, 50, and 100 (left to right). An auxiliary point of choosing a particular number of clumps from each galaxy is that we must remove galaxies with fewer than that number of clumps from the comparison. After the matching process, the samples compared in Figure \ref{fig:fil_sizeflux} have sizes of 42 (F) and 43 (NF) for $n_{\rm clumps}\geq 10$, 38 (F) and 33 (NF) for $n_{\rm clumps}\geq 25$, 20 (F) and 23 (NF) for $n_{\rm clumps}\geq 50$, and 28 (F) and 10 (NF) for $n_{\rm clumps}\geq 100$. To be clear, samples with $n_{\rm clumps}\geq 10$ also includes galaxies with at least 25, 50, and 100 clumps, etc. 

We find no statistically significant (or visually apparent) difference between clump sizes of our filament and non-filament galaxies for the cases of $n_{\rm clumps}\geq 10$, $n_{\rm clumps}\geq 25$, and $n_{\rm clumps}\geq 50$. In the case of $n_{\rm clumps}\geq 100$, the difference appears at face-value to be extremely statistically significant, with filament galaxies generally having smaller clumps than non-filament galaxies. In fact, a bootstrap test with the $n_{\rm clumps}\geq 100$ populations finds the same result. However, as we show below, this difference is likely caused by residual differences in the physical resolutions of the populations, which exist because our non-filament population is very small (10 galaxies).

The most important caveat when considering the strong difference in the $n_{\rm clumps}\geq 100$ case is the very small number of galaxies in the case of non-filament environments (10). With such small numbers, the PSF-matching algorithm does not work well (too few bins to minimise differences). We find that the median angular resolution for filament galaxies in this subsample is $\sim 0.06$~dex sharper (lower) than the median for non-filament galaxies. In addition, the median distance for filament galaxies is $\sim 0.07$~dex closer than for non-filament galaxies. 

In Section \ref{subsec:effects}, we quantified the effects of distance and angular resolution on clump sizes. For distance, the effect is a simple linear equation: $\log (\rm{average~size}_{\rm new}/{\rm average~size}_{\rm old}) =1.79\log (d_{\rm new}/d_{\rm old})$. For angular resolution, the effect is a quadratic function of the logarithmic change in FWHM and a (non-analytic) dependence on the original FWHM. We can approximate $\log (\rm{average~size}_{\rm new}/{\rm average~size}_{\rm old})$ as a 2-D interpolation function of the original FWHM and $\log ({\rm FWHM}_{\rm new}/{\rm FWHM}_{\rm old})$. For galaxies that are closer or have better angular resolution than the reference, we can simply use the relation with the opposite sign, though the assumption of symmetry is possibly flawed in the case of angular resolution. 

With these equations, we can estimate the sizes all clumps in the galaxy samples would have at a reference distance and FWHM. By effectively placing all clumps at the same observing conditions, we can try to erase the difficulty caused by the imperfect physical FWHM matching given the small number statistics, though the approximations do not take into account the scatter in the relations and assume symmetry between worsening and improving angular resolution. In any case, we place all clumps in the $n_{\rm clumps}\geq 100$ samples at a distance of 25 Mpc and an angular resolution of FWHM$=1.5\arcsec$. These values are approximately the medians of the galaxies' observing conditions, leading to the minimum possible shift in clump sizes and therefore the minimum possible influence of galaxy variance from the equations used. We find that, in fact, the modified clump distributions are nearly indistinguishable ($p=0.11$ even considering thousands of clumps). Therefore, we do not have conclusive evidence of a difference in the clump size distributions of filament and non-filament galaxies. A larger sample of clumpy filament and non-filament galaxies, ideally well matched in physical resolution, will be necessary for a more definitive statement.


\subsection{Filament galaxies have a slight tendency toward more peripheral clumps than non-filament galaxies.} \label{subsec:clumpdisp}

One of the primary tasks we can perform with resolved \Ha observations is mapping the positions where star formation is occurring. In this work, we tabulate the displacement between flux-weighted centroids of each clump and the flux-weighted centre of the galaxy. Once again, the galaxy is defined to be the region of each image within the convex hull of the points where the fourth Scarlet scale is over a threshold (almost always the 16th percentile of the fourth-scale decomposition). Since the galaxies in the sample have a vast range of sizes, to compare the clump distributions, we divide the radial distance of each clump to the centre by the measured semi-major axis in the \textit{r}-band. The semi-major axes have been measured using elliptical photometry with second-moment maps \citep[\texttt{kinemetry},][]{Krajnovic2006}. As stated earlier, they come from SGA-2020 \citep{MoustakasSGA2023} based on the Legacy Imaging Surveys \citep{DeyLegacy2019}. The ratio of clump displacement to semi-major axis is a unitless quantity that represents the relative location of the clumps.

\begin{figure*}
    \centering
    \resizebox{\hsize}{!}{
    \includegraphics{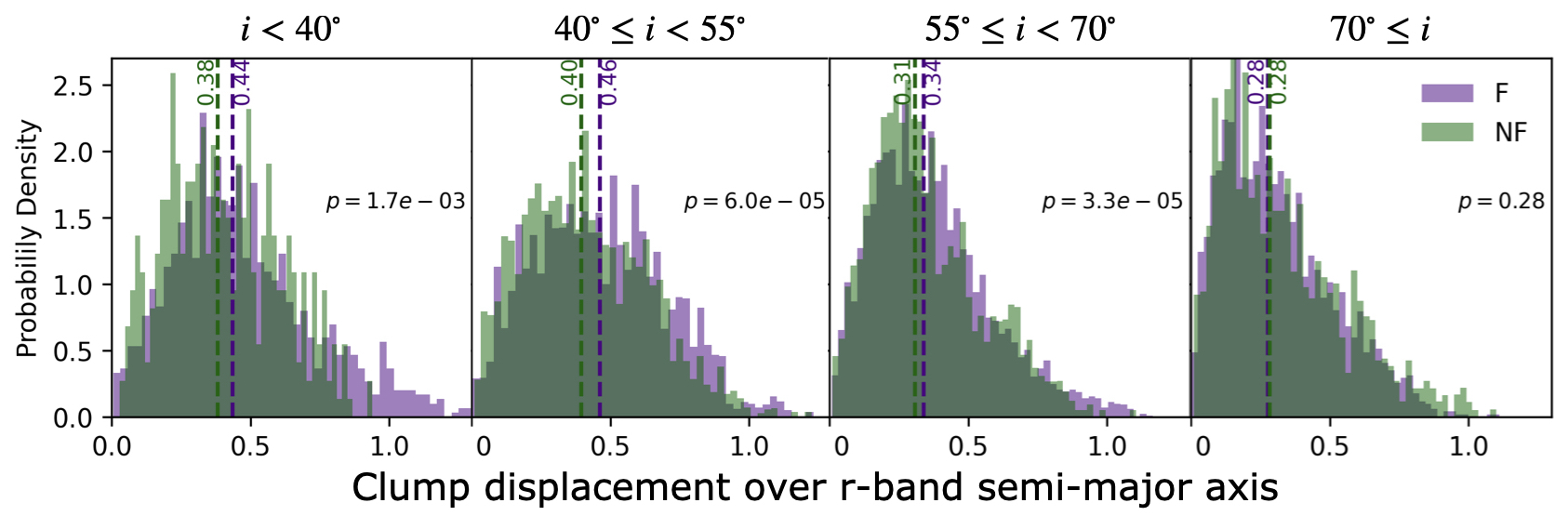}}
    \caption{Observed displacement (normalised by \textit{r}-band semi-major axis) of clumps in filament (purple) vs non-filament (green) galaxies in four bands of inclination, with medians shown as dashed lines. Between $0^\circ-70^\circ$, we find a small but statistically significant trend for filament galaxies to have more peripheral clumps.}
    \label{fig:fnfdarat3}
\end{figure*}

The distribution of clump displacements from a galaxy's centre is strongly dependent on its inclination as clumps along our line of sight can get merged, and for disky galaxies this is much more pronounced for edge-on than face-on configurations. For this reason, we compare only galaxies with similar inclinations. In Figure~\ref{fig:fnfdarat3}, we compare the distributions of observed clump displacement (normalised by semi-major axis) in filament (F; purple) and non-filament (NF; green) galaxies. We divide galaxies into four bins of inclination (0-40$^\circ$, 40-55$^\circ$, 55-70$^\circ$, and 70-90$^\circ$) and perform the FWHM-matching process (25 times) independently in each bin to suppress observational biases. After the matching process, the samples compared in Figure \ref{fig:fnfdarat3} have sizes of 37 (F) and 18 (NF) for 0-40$^\circ$, 13 (F) and 29 (NF) for 40-55$^\circ$, 46 (F) and 43 (NF) for 55-70$^\circ$, and 38 (F) and 23 (NF) for 70-90$^\circ$. (Not all galaxies in our \Ha sample have inclination measurements.)

The distributions peak around $r/a\sim 0.2-0.4$, with the lowest values in nearly edge-on galaxies (where the 3D distance from the centre is most underestimated because of projection issues). We see long tails toward larger values, including peripheral clumps outside the nominal stellar radius. In all galaxies except for nearly edge-on ones, we see a small but statistically significant trend for filament galaxies to have more outwardly/peripherally placed clumps than non-filament galaxies ($p=1.7 \times 10^{-3} - 2.3 \times 10^{-5}$). Bootstrap experiments with the galaxy samples (where we randomly select samples with the same length as the original but with replacement) also yield the same result, suggesting that the trend is quite stable. To understand the underlying processes causing this trend, we need to build simulations with spatially resolved star formation and multiphase ISM \citep[e.g.,][]{Moreno2021,Verwilghen2024} on the large scales of the cosmic web.


In Section \ref{subsec:clumpsmall}, we chose a certain number of clumps from each galaxy via a random selection in order to avoid biasing results toward the galaxies with the most clumps. This restricted the galaxies to those with at least the number of clumps chosen (10, 25, 50, and 100). For the displacement result, we do not adopt the same practice for two reasons: 1) When doing both the inclination bins and the clump number procedure, there are too few galaxies for comparison. 2) In the case of displacement, we are trying to understand how the clumps trace the \Ha distribution in the galaxy, not measure properties of individual clumps. Thus, it makes sense to include all clumps of each galaxy in the computation. Choosing, for example, just 10 clumps from a galaxy with 500 clumps cannot properly sample the \Ha distribution. On the other hand, it is much easier to sample the size distribution with fewer clumps. 


Nevertheless, when we do restrict the sample with the number of clumps (and do not bin by inclination), we find that the relation in Figure \ref{fig:fnfdarat3} is not limited to a few $n_{\rm clumps}\geq 100$ galaxies like in the size case. Rather, we find that it holds for the $n_{\rm clumps}\geq 50$, $n_{\rm clumps}\geq 25$, and $n_{\rm clumps}\geq 10$ cases. Thus, the danger of biasing toward galaxies with the most clumps is not an issue for the displacement trend.


However, we must caution that the optical size measurements of SGA-2020 are uncertain, and that a more robust conclusion will require SGA-2026 elliptical fits (Moustakas et.~al, in prep). Nevertheless, there is no reason for a systematic bias toward filament or non-filament galaxies. Also, the moment-based semi-major axis is a relatively unbiased tracer of galaxy size. In comparison, for example, a surface-brightness-defined axis depends strongly on the image observational depth, which we find in our case to be slightly different between filament and non-filament populations.

\section{Clumps Observed in Fractal Patterns}\label{sec:fractal}

In Section \ref{subsec:effects}, we discovered a strong power-law correlation between the number of clumps and distance ($n_{\rm cl} \propto d^{-1.35}$) in our experiment with artificially moving galaxies to larger distances. Due to the lack of an associated scale in a power law, this result suggests that \Ha clumps are self-similar structures. In other words, in our observations they behave as fractals would, with well-defined sizes depending on the zoom factor. This signifies that \HII regions are not randomly distributed in a galaxy but are rather organised in hierarchical structures. The clumps we observe are higher levels of that hierarchy and depend on the observing conditions. Thus, as long as we carefully match populations with their physical resolutions (Section \ref{subsec:fwhm}), we should be making a physically meaningful comparison.


The most fundamental property of a fractal is its dimension: If the mass of a clump at scale $R$ is $M$, we can define the fractal dimension $D$ such that $(M/M_0) \propto (R/R_0)^D$ \citep{Mandelbrot1983,PfennigerCombes1994}. While the physical implications are complex---fractal behaviour can be driven by various things, like turbulence, gravity, chemistry, etc.---by comparing our fractal dimension to those measured for other types of clouds, we can discover connections between different phenomena. With our distance experiment, we can estimate the value of the fractal dimension of \Ha clumps. Based on the assumptions of true self-similarity (consistent behaviour across all distance scales) and negligible inter-clump gas mass, we can use Equation 9 from \cite{PfennigerCombes1994} to show that the absolute value of the slope of the best-fit line in Figure \ref{fig:avgclump} is equal to the fractal dimension. In other words, our distance experiment suggests that $D\sim 1.3-1.4$. 

\cite{CaicedoOrtiz2015} use a similar method (measuring the log-log slope of the number of boxes to cover \Ha emission of a source vs.~the box size) to find the fractal dimension of two giant \HII regions and find $D\sim 1.4$ in both cases. The consistency in the fractal dimension of their \HII regions and the larger \Ha clumps in our study provide more evidence of the well-defined hierarchical organisation of \Ha emission from scales even smaller than individual clouds to large complexes of clouds. 

In addition, \cite{Elmegreen2006} measure the number of clumps as a function of angular resolution for galaxy NGC 628 and find a power-law slope of $\sim -1.5$, suggesting $D\sim 1.5$, similar to our value. Furthermore, our estimated fractal dimension is the same as that of isophotes of atomic and molecular gas clouds in the Milky Way and nearby galaxies from scales of 30 AU to a few hundred pc \citep[$D=1.3-1.4$, e.g.,][]{Scalo1990,Falgarone1992}. These papers determine the fractal dimension through a different method: measuring the correlation between area and perimeters of clouds, with isophotal boundaries. In our work, clump boundaries come from Scarlet wavelet decomposition, so we cannot make a direct comparison. Nevertheless, in all the aforementioned cases we are measuring fractal dimensions on 2-D projected images, so getting the same result perhaps reflects the connection between star formation and gas clouds.



One important consideration is that this simple calculation comes with assumptions and may not reflect the true fractal dimension. \cite{Elmegreensquared2001}, for example, take into account overlap and blending of observed structures (from projection effects, for example) in experiments with numbers of clumps as a function of PSF (modeled with Gaussian smoothing) by creating models of different fractal dimensions and comparing the resulting projected maps to observations of real galaxies. With similar slopes as ours (compare our Figure \ref{fig:psfexpsings} to their Figure 2), they find that the galaxies are most similar to the $D=2.3$ fractal model, rather than the $D=1.3$ model.

\section{Conclusions} \label{sec:conc}


Using resolved \Ha observations of a sample of 685 galaxies in and near the filaments around the Virgo cluster, we have systematically analysed the role of filaments on the morphology of star formation at small scales systematically. We have also employed the wealth of integrated data available for our galaxies, including stellar masses, morphologies, distances, and environmental information. We have developed a pipeline to identify and quantify star formation clumps in the \Ha images using wavelet decomposition (Scarlet) and image deblending (Photutils), respectively. Based on stringent visual checks, we reduced the 685 galaxies to a pure sample of 414 galaxies. When comparing clumps of galaxies from different populations, steps must be taken to ensure that differences are physical and not simply the result of observational biases. We have shown that the primary variable determining observed properties of the clump distribution is the physical-scale FWHM, so we have designed an algorithm to match FWHM distributions for unbiased comparisons. In the end, we match populations of 157 filament and 133 non-filament galaxies, for a total of 290 galaxies.

We find that in general, filament and non-filament galaxies have the same distribution of clump sizes. However, in terms of the radial distribution of clumps, for all galaxies except nearly edge-on ones ($i>70^\circ$), filament galaxies tend to have slightly more outer clumps than non-filament galaxies with high statistical significance ($\sim 4\sigma$). Our clump displacement measurements are normalised by stellar size measurements from SGA-2020 \citep{MoustakasSGA2023}, which will come with improvements in SGA-2026. These will allow us to make more robust statements. Further investigation, including simulation work, will be needed to understand the origin of this result.

Through our tests studying the effects of distance and angular resolution on clump properties, we have shown that \Ha clumps are fractal/hierarchical in nature, not random/homogeneous, consistent with studies of both \HII regions \citep[e.g.,][]{CaicedoOrtiz2015} and atomic and molecular interstellar gas \citep[e.g.,][]{Falgarone1992}. Moving toward comparison of filament and non-filament environments, filament galaxies are clearly in regions of higher local density than their non-filament counterparts. In our sample, they tend to have slightly smaller sizes (both recent-SF/\Ha and stellar/optical footprints), though not in highly statistically significant trends.

In general, repeating the analysis presented in this work on higher-angular-resolution \Ha images is required to make stronger statements about differences in the clump distributions. For example, the Javalambre Photometric Local Universe Survey \citep[J-PLUS;][]{CenarroJavalambre2014,CenarroJplus2019} delivers better and much more consistent seeing \citep[$1.1$\arcsec in the \textit{r}-band;][]{RahnaHa2025} than what is available for this work. The J-PLUS Northern Galactic Hemisphere footprint covers ten out of the thirteen identified Virgo filaments, with seven having partial or full coverage within the two priority areas. Furthermore, one of the J-PLUS filters (J0660) contains the \Ha emission line up to $z<0.017$, thus allowing us to measure resolved \Ha fluxes for all Virgo filament galaxies and their nearby non-filament counterparts in the footprint.

The role of the more tenuous and large-scale higher-density environments in filaments in affecting star formation has not yet been studied extensively, especially compared to clusters. This study creates a technical framework to evaluate resolved star formation and applies it to a medium-size sample (414 galaxies) to study systematic effects of filaments. However, our understanding is far from complete. We will need both detailed simulations and much larger galaxy samples with better angular resolution as described above. Spectroscopy would allow us to pinpoint details of not only the youngest stellar populations (with \Ha, for example) but also older populations through features like the Balmer break and absorption lines to probe the timescales on which filaments influence galaxies. 

Several complementary studies are already near completion. For example, an analysis of very high resolution and sensitive MeerKAT \HI spectral cubes of galaxies along the Virgo III filament as well as an investigation of the molecular gas in several \HI-deficient galaxies with ALMA, NOEMA, and SMA will help illuminate the effects of filaments on the gas components of galaxies. By combining comprehensive integrated properties of galaxies in a wide range of local densities around filaments with resolved observations of \Ha, CO, and \HI, we will be able to better understand the range of physical processes operating in the cosmic web.

\begin{acknowledgements}
This work was supported by the Swiss National Science Foundation (SNSF) under funding reference 200021\_213076 ``Galaxy evolution in the cosmic web".
This research has made use of the Astrophysics Data System, funded by NASA under Cooperative Agreement 80NSSC21M00561. This research has made use of the NASA/IPAC Extragalactic Database, which is funded by the National Aeronautics and Space Administration and operated by the California Institute of Technology. YMB acknowledges support from UK Research and Innovation through a Future Leaders Fellowship (grant agreement MR/X035166/1).  GHR acknowledges the support of NSF AAG grants 1716690 and 2308126 as well as NASA ADAP grant 80NSSC21K0641. This research was supported by the International Space Science Institute (ISSI) in Bern and the Institute for Fundamental Physics of the Universe (IFPU) in Trieste.

\end{acknowledgements}

\bibliographystyle{aa} 
\bibliography{Halpha} 

\end{document}